\documentclass[useAMS,usenatbib]{mn2e}
\usepackage{graphicx}
\usepackage[dvips]{color}
\usepackage{txfonts}
\usepackage{amsfonts}
\usepackage{amssymb}
\usepackage{natbib}
\usepackage{multirow}

\newcommand{\te}{$T_{\rm eff}$}
\newcommand{\logg}{$\log{g}$}
\newcommand{\vsini}{$v\sin{i}$}
\newcommand{\Veq}{\ensuremath{V_{\rm eq}}}
\newcommand{\Vsync}{\ensuremath{V_{\rm synch}}}
\newcommand{\kms}{km\,s$^{-1}$}
\newcommand{\cd}{d$^{-1}$}
\newcommand{\mhz}{$\mu$Hz}
\newcommand{\mc}[1]{\multicolumn{2}{c}{#1}}

\title[V380\,Cyg: orbital solution and intrinsic variability of
the primary]{The eccentric massive binary V380\,Cyg: revised orbital
elements and interpretation of the intrinsic variability of the
primary component\thanks{Based on the data gathered with NASA's
Discovery mission
    {\it Kepler} and with the HERMES spectrograph, installed at the Mercator
    Telescope, operated on the island of La Palma by the Flemish Community, at
    the Spanish Observatorio del Roque de los Muchachos of the Instituto de
    Astrof\'{\i}sica de Canarias and supported by the Fund for Scientific
    Research of Flanders (FWO), Belgium, the Research Council of K.U.Leuven,
    Belgium, the Fonds National de la Recherche Scientific (F.R.S.--FNRS),
    Belgium, the Royal Observatory of Belgium, the Observatoire de Gen\`eve,
    Switzerland and the Th\"uringer Landessternwarte Tautenburg, Germany.}}

\author[A. Tkachenko et al.]{A.\ Tkachenko,$^{1,}$\thanks{Postdoctoral Fellow of the Fund for Scientific Research (FWO), Flanders, Belgium} P.\ Degroote,$^{1,}$\thanks{Postdoctoral Fellow of the Fund for Scientific Research (FWO), Flanders, Belgium} C.\ Aerts,$^{1,2}$ K.\ Pavlovski,$^3$
  J.\ Southworth,$^4$ \newauthor P.~\ I.\ P\'{a}pics,$^1$ E.\ Moravveji,$^1$ V.\ Kolbas,$^3$ V.\ Tsymbal,$^5$ J.\ Debosscher,$^1$ and K.\ Cl\'{e}mer$^1$\\
  $^1$Instituut voor Sterrenkunde, KU Leuven, Celestijnenlaan 200D, B-3001 Leuven, Belgium\\
  $^2$Department of Astrophysics, IMAPP, Radboud University Nijmegen, 6500 GL Nijmegen, The Netherlands\\
  $^3$Department of Physics, University of Zagreb, Bijeni\v{c}ka cesta 32, 10000 Zagreb, Croatia\\
  $^4$Astrophysics Group, Keele University, Staffordshire ST5 5BG, UK\\
  $^5$Tavrian National University, Department of Astronomy, Simferopol, Ukraine}

\date{Received date; accepted date}

\pagerange{\pageref{firstpage}--\pageref{lastpage}} \pubyear{2013}

\begin{document}

\label{firstpage}

\maketitle

\begin{abstract}
  We present a detailed analysis and interpretation of the high-mass binary V380
Cyg, based on high-precision space photometry gathered with the {\it
Kepler} space mission as well as high-resolution ground-based
spectroscopy obtained with the {\sc hermes} spectrograph attached to
the 1.2m Mercator telescope.  We derive a precise orbital solution
and the full physical properties of the system, including dynamical
component mass estimates of 11.43$\pm$0.19 and
7.00$\pm$0.14~M$_\odot$ for the primary and secondary, respectively.
Our frequency analysis reveals the rotation frequency of the primary
in both the photometric and spectroscopic data and additional
low-amplitude stochastic variability at low frequency in the space
photometry with characteristics that are compatible with recent
theoretical predictions for gravity-mode oscillations excited either
by the convective core or by sub-surface convective layers. Doppler
Imaging analysis of the Silicon lines of the primary suggests the
presence of two high-contrast stellar surface abundance spots which
are located either at the same latitude or longitude. Comparison of
the observed properties of the binary with present-day single-star
evolutionary models shows that the latter are inadequate and lack a
serious amount of near-core mixing.
\end{abstract}


\section{Introduction}

V380\,Cyg (HR\,7567, HD\,187879, KIC\,5385723) is a bright
($V$=5.68) double-lined spectroscopic binary (SB2) consisting of two
B-type stars residing in an eccentric 12.4~d orbit. The
primary component is an evolved star which is about to start its
hydrogen shell burning phase, whereas the secondary, fainter
component is a main-sequence star comparable in temperature. Most of
the system's brightness comes from the primary, with the secondary
contributing about 6\% of the total flux. For this reason, the
system was thought to be a single-lined spectroscopic binary for a
long time until \citet{Batten1962} reported the detection of the
secondary's spectral lines on twenty two newly obtained
spectrograms. Although the double-lined nature of the star was a
subject of debate for another twenty years \citep[see
e.g.,][]{Popper1981}, \citet{Hill1984} confirm that V380\,Cyg is an
SB2 star based on newly obtained high dispersion spectroscopic data.
In addition to the orbital elements which differ insignificantly
from those derived by \citet{Batten1962}, the authors report on
apsidal motion of about 0.011 degrees/period.

Since then, V380\,Cyg was the subject of several more extensive
studies based on both ground-based spectroscopy and multicolour
photometry. \citet[][hereafter G2000]{Guinan2000} combined newly
obtained differential \emph{U,B,V} photometry with the spectroscopic
data obtained by \citet{Popper1998} to refine the orbital solution
and to determine precise physical properties of both binary
components. The times of light minima were used to measure apsidal
motion of the system which was found to be of 24 degrees per 100
years (0.008 degrees/period), slightly lower than but in reasonable
agreement with the value reported by \citet{Batten1962}.

\begin{table}
\tabcolsep 1.5mm\caption{Physical parameters of both stellar
components of the V380\,Cyg binary system as derived by
\citet{Guinan2000} and \citet{Pavlovski2009}. Parameter errors are
given in parenthesis in terms of last digits.}\label{Table1}
\begin{tabular}{lr@{.}lr@{.}lr@{.}lr@{.}l} \hline
Parameter & \multicolumn{4}{c}{Primary} &
\multicolumn{4}{c}{Secondary}\\
 & \multicolumn{2}{c}{G2000} & \multicolumn{2}{c}{P2009} & \multicolumn{2}{c}{G2000} & \multicolumn{2}{c}{P2009} \\\hline
Mass (M$_{\odot}$) & 11&10(50) & 13&13(24) & 6&95(25) & 7&78(10)\\
Radius (R$_{\odot}$) & 14&70(20) & 16&22(26) & 3&74(07) & 4&06(08)\\
$\log$ (L/L$_{\odot}$) & 4&60(03) & 4&73(03) & 3&35(04) & 3&51(04)\\
\te\ (K) & \multicolumn{2}{l}{21\,350(400)} & \multicolumn{2}{l}{21\,750(280)} & \multicolumn{2}{l}{20\,500(500)} & \multicolumn{2}{l}{21\,600(550)}\\
\logg\ (dex) & 3&148(023) & 3&136(014) & 4&133(023) & 4&112(017)\\
M$_{\rm bol}$ (mag) & -6&75(07) & -7&06(06) & -3&62(10) & -4&03(10)\\
V$_{\rm eq}$ (\kms) & \multicolumn{2}{l}{98(4)} & \multicolumn{2}{l}{98(2)} & \multicolumn{2}{l}{32(6)} & \multicolumn{2}{l}{43(4)}\\
\hline
\end{tabular}
\end{table}

\begin{table}
\tabcolsep 2.5mm\caption{List of the spectroscopic observations of
V380\,Cyg. JD is the Julian Date, exposure time stands for the mean
value for each run, and $N$ gives the number of spectra taken during
the corresponding period.}\label{Table2}
\begin{tabular}{llr@{.}lr} \hline
\multicolumn{2}{c}{Time period} & \multicolumn{2}{c}{Exposure time} & $N$\\
\multicolumn{1}{c}{Calendar date} &
\multicolumn{1}{c}{JD-2\,450\,000} & \multicolumn{2}{c}{sec} &
\\\hline
05.04--07.04.2011 & 5656.72--5658.76 & 952&67 & 7\\
09.04--16.04.2011 & 5660.65--5667.74 & 894&28 & 45\\
19.04--20.04.2011 & 5670.64--5671.70 & 853&85 & 13\\
03.05--16.05.2011 & 5684.59--5697.70 & 1100&00 & 62\\
21.05--25.05.2011 & 5702.55--5706.58 & 879&73 & 22\\
28.05--30.05.2011 & 5710.49--5711.69 & 900&00 & 14\\
02.06--06.06.2011 & 5714.53--5718.72 & 1200&00 & 21\\
12.06--13.06.2011 & 5724.52--5725.72 & 1200&00 & 20\\
02.07--03.07.2011 & 5744.55--5744.74 & 1200&00 & 14\\
06.07--07.07.2011 & 5749.38--5749.74 & 1380&95 & 21\\
15.07--26.07.2011 & 5758.41--5769.41 & 1199&73 & 72\\
31.07--01.08.2011 & 5774.39--5774.74 & 1200&00 & 23\\
08.08--09.08.2011 & 5782.39--5782.67 & 1200&00 & 20\\
20.08--25.08.2011 & 5794.35--5798.66 & 1200&00 & 52\\
02.08--04.08.2012 & 6142.40--6143.71 & 1223&48 & 46\\\cline{5-5}
\multicolumn{4}{l}{Total number of spectra:\rule{0pt}{11pt}} & {\bf 452}\\
\hline
\end{tabular}
\end{table}

Eclipsing binary stars are a prime source for testing evolutionary
models, as their fundamental parameters like masses, radii, etc. can
be derived to a very high precision. The proper solution of the
convective energy transport is one of the biggest issues of the
current evolutionary models. However, comparison of the observed
stellar properties with the theoretical predictions allows the
deduction of an important parameter such as the convective
overshooting $\alpha_{ov}$ (that stands for the stellar convective
core over-extension). For V380\,Cyg, G2000 found that the position
of the more massive primary on the Hertzsprung-Russell (HR) diagram
requires an extremely high value of $\alpha_{ov}$ of about
0.6$\pm$0.1 to match the observed properties of the star with the
theoretical models. This is actually the highest known value of
$\alpha_{ov}$ for B-type stars, whereas the typical value is of the
order of 0.2 for single stars with similar masses \citep[see
e.g.,][]{Aerts2003,Aerts2011,Briquet2011}. Assuming this typical
value of $\alpha_{ov}$ for the primary of V380\,Cyg leads to
inconsistency between the mass deduced from the binary dynamics and
the one obtained from the evolutionary models. This inconsistency
between the two masses is usually called the {\it mass discrepancy}.
It is worth noting that the second highest value after V380\,Cyg of
$\alpha_{ov}\sim$0.45, has also been found for the
$\sim$8~M$_{\odot}$ primary component of a detached binary system
\citep{Briquet2007}. Alternatively, the rate of the apsidal motion,
depending on the internal mass distribution of a star, can be used
to estimate the overshooting parameter $\alpha_{ov}$. The value
obtained by G2000 in this way perfectly agrees with the one deduced
from the position of the star in the HR diagram. On the other hand,
\citet{Claret2007} found that the apsidal motion rate of the
V380\,Cyg is compatible with $\alpha_{ov}=$0.4$^{+0.2}_{-0.3}$.

\citet[][hereafter P2009]{Pavlovski2009} collected about 150
high-resolution \'{e}chelle spectra using several telescopes and
performed a detailed spectroscopic study of the system with the aim
to determine the precise orbital solution as well as the physical
parameters of both stellar components. The authors also revisited
the \emph{U,B,V} light curves obtained by G2000 and analysed the
data using a modified version of the Wilson-Devinney code
\citep{Wilson1971,Wilson1979,Wilson1993}. The spectral disentangling
(SPD) technique applied by P2009 to their spectra, besides providing
precise orbital elements, revealed a faint secondary component in
the composite spectra and provided high-quality disentangled spectra
for both stars. The spectra were further analysed to determine the
fundamental parameters as well as to do the abundance analysis for
the primary. We refer to Table~\ref{Table1} for an intercomparison
between the physical parameters of the stars derived by G2000 and
P2009.

Similar to G2000, \citet{Pavlovski2009} found that the position of
the more massive primary in the HR diagram does not match the
theoretical predictions for a single star of a similar mass. The
authors find a mass discrepancy of about 1.5~M$_{\odot}$ for the
primary component of V380\,Cyg. Though this is much less than what
was found for V380\,Cyg by, e.g., \citet{Lyubimkov1996} who reported
a mass discrepancy of 3.4 and 1.1~M$_{\odot}$ for the primary and
the secondary, respectively, the value is still significant. The
mass discrepancy problem observed in massive O- and B-type stars has
been known for more than twenty years already and has been discussed
in detail by \citet{Herrero1992}. \citet{Hilditch2004} pointed out
that the discrepancy does not disappear when the effects of rotation
are included into the models, P2009 came to the same conclusion
concerning the primary of V380\,Cyg.

It is thus important to have an independent estimate of the core
overshooting parameter $\alpha_{ov}$ for this type of stars, and
this is provided by the asteroseismic methods. In \citet[][hereafter
Paper~I]{Tkachenko2012}, we presented results of the preliminary
analysis of our data consisting of high-quality {\em Kepler} space
photometry and high-resolution, ground-based spectroscopy. We
confirmed the double-lined nature of the binary and presented
precise orbital elements, in general agreement with the two previous
findings by G2000 and P2009. Both photometric and spectroscopic data
revealed the presence of variability intrinsic to the primary
component, which we interpreted in terms of stochastically excited
gravity (g-) mode oscillations. Moreover, both data sets showed that
at least part of this variability occurs at (higher) harmonics of
the orbital frequency suggesting a tidal nature for the pulsations.

In this paper, we present a solid interpretation of our full data
set that consists of the data presented in Paper~I and additional
four quarters of {\em Kepler} data and some 50 high-dispersion, high
signal-to-noise ratio (S/N) spectra. The data themselves as well as the
data reduction procedure are described in Section~2. Photometric
analysis of the system is presented in Section~3 while Section~4 is
devoted to the analysis of our spectra. Discussion and conclusions
are presented in Section~5.

\section{Observations and data reduction}

The photometric analysis of the V380\,Cyg system presented in this
paper is based on the high-quality {\em Kepler} space data. The data
are released in quarters (periods of time between the two
consecutive spacecraft rolls needed to keep the solar shutters
pointing towards the Sun), each quarter comprises about three months
of nearly continuous observations. The telescope operates in two
different modes, long- and short-cadence (LC and SC, respectively),
which differ from each other by the time of integration which is
29.42~min for the LC and 58.85~s for the SC data. The exposure time
is the same in both modes and equals 6.54~s (0.52~s of readout
included). Thus, the two modes basically differ by the
time-resolution where the LC integrates over 270 single exposures
whereas one data point in the SC mode contains 9 exposures. For
V380\,Cyg, we use all available {\em Kepler} data, i.e. Q7, Q9-Q10,
and Q12-Q14, which gives a time span of $\sim$560 days.

\begin{figure}
\includegraphics[scale=0.88]{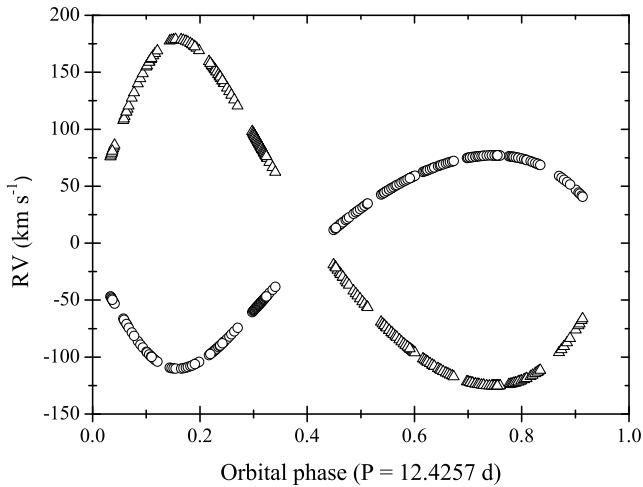}
\caption{\label{Figure0} Orbital phase distribution of
out-of-eclipse spectra in terms of the radial velocities (RV)
computed from our final orbital solution (see Section~4.1, Table 4).
The primary and secondary RVs are shown by circles and triangles,
respectively. Phase zero corresponds to the periastron passage.}
\end{figure}

Because of the extreme brightness of the system ($V$=5.68), a
dedicated customized aperture mask was defined to collect the flux
from the star. This is fully justified given that {\em Kepler} has
proved already that it can provide high-precision photometry even
for highly saturated targets \citep[see
e.g,][]{Bryson2010,Kolenberg2011,Tkachenko2012}.

In addition to the photometric data, high-dispersion ground-based
spectroscopy has been acquired with the {\sc hermes} spectrograph
\citep[][]{Raskin2011} attached to the 1.2-metre Mercator telescope
(La Palma, Canary Islands). All spectra have a resolving power of
$\sim$85\,000 and cover the wavelength range from 380 to 900~nm.
Table~\ref{Table2} gives the journal of observations listing the
calendar and the Julian dates in its first two columns, and the mean
exposure time and the number of spectra obtained during each run in
columns three and four, respectively. In total, we have obtained 452
spectra with an average S/N above 200, of which 406 were part of the
analysis presented in Paper~I. More information about the
spectroscopic data and the reduction procedure can be found in
Paper~I. Figure~\ref{Figure0} shows the phase distribution of all
our out-of-eclipse spectra that will be analysed in Section~4. The
radial velocities (RV) presented in this Figure were computed from
our final orbital solution (cf. Table~4) and are used here for
illustration purposes.

\section{{\em Kepler} photometry}

In this section, we describe the procedure used to obtain a
photometric orbital solution as well as the results of the frequency
analysis of the residuals computed after subtracting our best fit
binary model from the original data.

\subsection{Orbital solution}

\begin{figure*}
\includegraphics[width=\textwidth,angle=0]{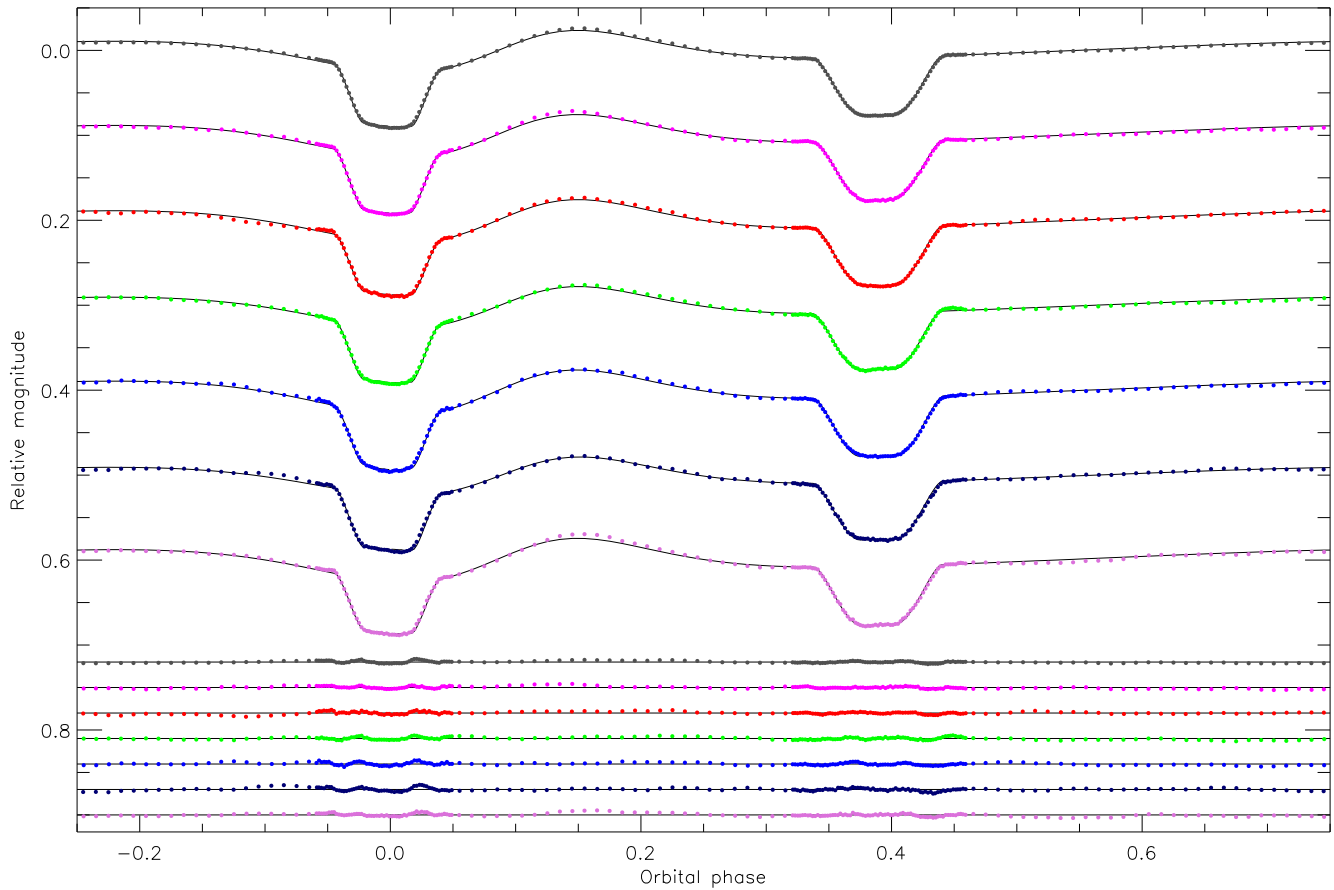}
\caption{\label{Figure1} Best fits of the {\it Kepler} satellite
light curves of V380\,Cyg using {\sc wd2004}. The binned data are
shown in the upper part of the plot and the residuals of the fits in
the lower part of the plot. The fits are shown for the combined data
and for individual quarters. The colour version of the figure is
available online only.}
\end{figure*}

\begin{table}
\caption{Summary of the parameters for the {\sc wd2004} solution of
the {\it Kepler} short-cadence light curve of V380\,Cyg. For further
details on the control parameters please see the {\sc wd2004} user
guide \citep{Wilson2004}. A and B refer to the primary and secondary
stars, respectively. The uncertainties on the photometric parameters
come from the scatter of appropriate solutions with different sets
of fitted parameters, and are much greater than the formal errors of
the best single solution.} \setlength{\tabcolsep}{4pt}
\begin{tabular}{llc} \hline
Parameter                           & {\sc wd2004} name     & {Value$\pm$Error}            \\
\hline
{\it Control and fixed parameters:} \\
{\sc wd2004} operation mode         & {\sc mode}            & 2                \\
Treatment of reflection             & {\sc mref}            & 2 (detailed)     \\
Number of reflections               & {\sc nref}            & 2                \\
Limb darkening law                  & {\sc ld}              & 1 (linear)       \\
Numerical grid size (normal)        & {\sc n1, n2}          & 60               \\
Numerical grid size (coarse)        & {\sc n1l, n2l}        & 60               \\[3pt]
{\it Fixed parameters:} \\
Orbital period (d)                  & {\sc period}          & 12.425719        \\
Primary eclipse time (HJD)          & {\sc hjd0}            & 2441256.544      \\
Mass ratio                          & {\sc rm}              & 0.6129           \\
$T_{\rm eff}$ star\,A (K)           & {\sc tavh}            & 21\,700          \\
Rotation rates                      & {\sc f1, f2}          & 1.5, 2.0         \\
Gravity darkening                   & {\sc gr1, gr2}        & 1.0              \\
Bolometric albedos                  & {\sc alb1, alb2}      & 1.0              \\
Bolometric LD coeff.\ A             & {\sc xbol1}           & 0.645            \\
Bolometric LD coeff.\ B             & {\sc xbol2}           & 0.681            \\
Third light                         & {\sc el3}             & 0.0              \\
Passband LD coeff.\ B               & {\sc x2}              & 0.224            \\
Orbital eccentricity                & {\sc e}               & 0.2224           \\[3pt]
{\it Fitted parameters:} \\
Phase shift                         & {\sc pshift}          & $-$0.0502$\pm$0.0003    \\
Star\,A potential                   & {\sc phsv}            &     4.676$\pm$0.020     \\
Star\,B potential                   & {\sc phsv}            &     11.06$\pm$0.11      \\
Orbital inclination (\degr)         & {\sc xincl}           &     80.76$\pm$0.24      \\
Longitude of periastron (\degr)     & {\sc perr0}           &     138.4$\pm$0.39      \\
$T_{\rm eff}$ star\,B (K)           & {\sc tavc}            &     23840$\pm$500       \\
Light from star\,A                  & {\sc hlum}            &    12.045$\pm$0.006     \\
Passband LD coeff.\ A               & {\sc x2}              &     0.137$\pm$0.064     \\[3pt]
{\it Derived parameters:} \\
Light from star\,B                  & {\sc clum}            &     0.750$\pm$0.012     \\
Fractional radius of star\,A        &                       &    0.2633$\pm$0.0015    \\
Fractional radius of star\,B        &                       &   0.06403$\pm$0.00072   \\[3pt]
{\it Physical properties:} \\
Mass of star\,A ($M_\odot$)         &                       &     11.43$\pm$0.19      \\
Mass of star\,B ($M_\odot$)         &                       &     7.00$\pm$0.14      \\
Radius of star\,A ($R_\odot$)       &                       &     15.71$\pm$0.13      \\
Radius of star\,B ($R_\odot$)       &                       &     3.819$\pm$0.048     \\
Surface gravity star\,A (log$g$)    &                       &     3.104$\pm$0.006     \\
Surface gravity star\,B (log$g$)    &                       &     4.120$\pm$0.011     \\
Orbital semimajor axis ($R_\odot$)  &                       &     59.65$\pm$0.35      \\
$\log(Luminosity A / L_\odot)$       &                       &     4.691$\pm$0.041     \\
$\log(Luminosity B / L_\odot)$       &                       &     3.626$\pm$0.038     \\
$M_{\rm bol}$ star\,A (mag)         &                       &   $-$6.98$\pm$0.10      \\
$M_{\rm bol}$ star\,B (mag)         &                       &   $-$4.31$\pm$0.10      \\
Distance (pc)                       &                       &       970$\pm$21        \\
\hline
\end{tabular}
\label{Table3}
\end{table}

The effects of binarity in the light curves were modelled using the
Wilson-Devinney (WD) code \citep{Wilson1971,Wilson1979}. We used the
{\sc jktwd} wrapper \citep{Southworth2011} to drive automatic
iterations of the 2004 version of the code (hereafter called {\sc
wd2004}).

The data were first reduced to a more manageable quantity by
phase-binning. This comprised converting the timestamps into orbital
phase using the ephemeris of G2000, sorting in phase, and then
binning them into 200 normal points. The bin sizes were chosen to be
five times smaller during the eclipses, in order to provide a finer
sampling of the light variations at these important times. This
binning process was performed for both the combined data and for the
six individual quarters, and was fundamental to making the problem
tractable. The WD code becomes very slow when considering either
orbital eccentricity or a high numerical accuracy; both are required
for the current data. Even after binning the data to 200 normal
points, it took 20--30 hours of CPU time to fit a single light
curve.

Our analysis in the current work was guided by that of the Q7 data
in Paper~I. Multiple alternative solutions were calculated, but our
final numbers rest on the following characteristics (see
Table~\ref{Table3}). The detailed treatment of reflection in {\sc
wd2004} was used, with two reflections, and the numerical grid size
set to 60 for both the coarse and the fine grids. The stars were
specified to rotate at 1.5 and 2 times their pseudo-synchronous
velocities. The gravity darkening coefficients and albedos were set
to unity and third light was fixed at zero. A linear limb darkening
law was specified, and the bolometric coefficients for both stars
plus the coefficient for star\,B were fixed to values obtained from
interpolation within the tables of \citet{VanHamme1993}.

The fits were performed in Mode 2, in which the effective
temperatures and passband-specific light contributions of the two
stars are coupled using the predictions from model atmospheres. We
found this to provide similar fitted parameters as for solutions in
Mode 0 (see Paper~I) but a significantly lower $\chi^2$. {\sc
wd2004} unfortunately does not include predictions for the {\it
Kepler} passband, so we calculated solutions for the Johnson $V$ and
$R$ and the Cousins $R$ passbands. We fixed the effective
temperature of star\,A at 21\,700\,K and fitted for the effective
temperature of star\,B. Whilst our final results yield an
uncertainty of roughly $500$\,K on the effective temperature of
star\,B, the choice of different passbands moves the best-fitting
value around by no more than 60\,K. We therefore used the Cousins
$R$ passband in our final solutions, and added an extra $\pm$100\,K
onto the uncertainties of the effective temperature of star\,B.

Our exploratory fits revealed that the light curve yields poorer
constraints on the orbital shape ($e$ and $\omega$) than expected.
The quantities $e$ and $\omega$ are strongly correlated, a common
situation when studying eclipsing binaries, making it possible to
match the light curve for $0.20 < e < 0.24$. The time difference
between adjacent primary and secondary eclipses puts an excellent
constraint on $e\cos\omega$, making this quantity well-determined,
whereas spectroscopy yields the quantity $e\sin\omega$ to a good
precision. Without the ability to fit directly for these quantities
in either {\sc wd2004} or in spectral disentangling, we instead
constrained the orbital shape by iterating between the two analyses.
We used the light curve to find the best $\omega$ for a given $e$,
and the disentangling to find the best $e$ for a given $\omega$.
This procedure converged quickly to a solution which well satisfied
both types of data, and the final solutions for the light curve
analysis were performed with $e$ fixed at $0.2224$.

For our adopted solution we fitted for a phase shift with respect to
the orbital ephemeris of G2000, the potentials of the two stars,
$\omega$, the orbital inclination,  the passband-specific light
contribution and linear limb darkening coefficient of star\,A, and
the effective temperature of star\,B. We obtained fits for the full
(binned) data as well as for the six quarters individually. Our
final values for the fitted parameters refer to the best fit to the
full data, and our errorbars are the standard deviation of the
values from the six individual quarters. These results are given in
Table~\ref{Table3} and the best fits are plotted in
Figure~\ref{Figure1}. We also give the fractional radii (stellar
radii divided by the orbital semimajor axis) which are needed to
calculate the physical properties of the two stars.

It might be justified to divide the errorbars of our final
parameters by $\sqrt{6}$, as we modelled the six quarters of {\it
Kepler} data separately, but we have refrained from doing so. This
more conservative approach reflects the reality of the situation:
the {\sc wd2004} code is not capable of fitting the data perfectly
(see Figure~\ref{Figure1}) and the dominant ``noise'' (i.e.\ signal
not arising from the effects of binarity) is due to the intrinsic
variability of the primary rather than to Poisson statistics. The
final parameters are nevertheless very well determined, and lead to
measurements of the masses and radii of the stars to precisions
approaching 1\%. Agreement with the parameters found by previous
studies (P2009 and Paper~I) is also good. The errorbars are
typically an order of magnitude larger than the formal errors
calculated by {\sc wd2004} from the covariance matrix, as expected
due to the intrinsic variability of the primary.


Table\,\ref{Table3} also contains the full physical properties of
the V380\,Cyg system, calculated using the {\sc absdim} code
\citep{Southworth2005} and our new spectroscopic and photometric
results. For the distance, we quote the value obtained using the
2MASS $K_s$ apparent magnitude, bolometric corrections from
\citet{Bessell1998}, and adopting a reddening of $E(B-V) = 0.21 \pm
0.03$\,mag.

\subsection{Frequency analysis}

The residuals obtained after the orbital fit subtraction were
subjected to further frequency analysis, performed on both the
combined data set and the data from individual quarters separately.
For the extraction of individual frequencies, amplitudes, and
phases, we used the Lomb-Scargle version of the discrete Fourier
transform \citep{Lomb1976,Scargle1982} and consecutive prewhitening.
A more detailed mathematical description of the procedure can be
found in \citet{Degroote2009}.

The highest amplitude ($\sim$1~mmag) contribution of
0.08054~\cd detected in the residuals of all data sets is
close to the orbital frequency f$_{\rm orb}=1/P=$0.08048~\cd. The
difference between the two values is insignificant given the
Rayleigh limit of 0.0013~\cd\ for our photometric data. We also
detected a series of higher harmonics of the orbital frequency with
amplitudes at least twice as low.

The second dominant frequency in the combined data set was found to
be of 0.12946~\cd\ (1.4978~\mhz). The analysis performed on the data
from individual quarters showed that this peak is variable in time,
both in frequency and amplitude. We also detected the second
harmonic at 0.25879~\cd\ (2.9942~\mhz) of this frequency, and
together with the orbital frequency, these are the three dominant
peaks in the combined data set (left part of
Table~\ref{Table:Frequency analysis}, labelled ``Photometry''). From
the determined radius, inclination angle, and projected rotational
velocity of the primary (cf. Table~\ref{Table3} and Section~4.2), we
estimate the true rotation frequency of the star to be
$\sim$0.1250~\cd\ (1.4463~\mhz). This value is slightly different
from 0.12946~\cd\ (1.4978~\mhz) deduced from the observed light
variations, which is not unexpected given that the former value
represents the rotational frequency averaged over the stellar
interior. We thus speculate that the frequency of 0.12946~\cd\ is
associated with the rotationally modulated signal originating from
the primary component of V380\,Cyg. The signal itself was found to
be slightly non-sinusoidal, and the higher frequency value (compared
to the true rotation frequency) might be evidence for differential
rotation (in the sense that the spot rotates faster than the bulk of
the star). We explore the possibility of observing the spot signal
in more detail in Section~4.5.

\begin{figure}
\includegraphics[scale=0.48]{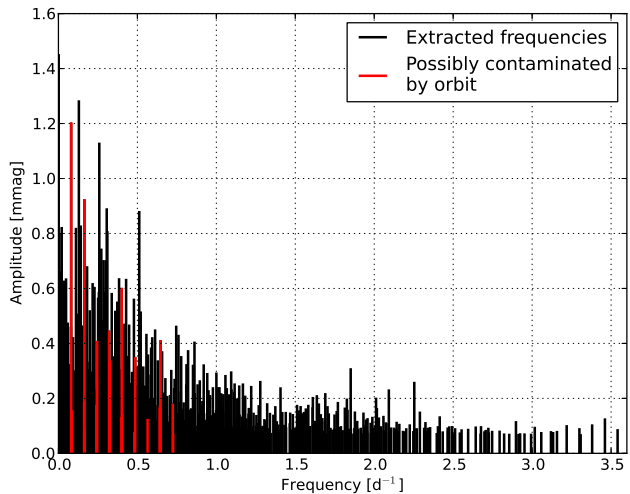}
\caption{\label{Figure2} Results of the frequency analysis of the
photometric residual signal after subtraction of the orbital
solution from the combined data set. Thick red lines indicate the
peaks possibly contaminated by the orbit.}
\end{figure}

In total, we detected about 300 significant frequencies in the
combined data set assuming a S/N of 4.0 as a stop criterion
\citep{Breger1993}. Figure~\ref{Figure2} shows all significant
frequencies and their amplitudes extracted from the combined data
set; those possibly contaminated by the binary orbit are
indicated by the thick red lines. We can see that the frequency
density is rather high and that there is a clear trend of decreasing
amplitude as the frequency increases. The time-frequency analysis
performed on individual quarters showed that the signal is variable
both in appearance and amplitude, suggesting its stochastic nature.
This confirms our earlier findings presented in Paper~I.

\subsection{Interpretation of the intrinsic variability of the primary}

\begin{figure}
\includegraphics[scale=0.47]{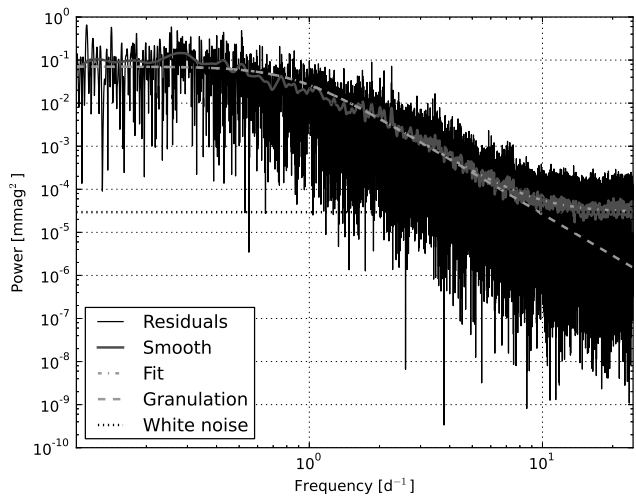}
\caption{\label{Figure3} Power spectrum (black) of the orbital
residual signal of V380\,Cyg with the fit by a granulation (grey
dashed line) and a granulation + white noise (grey dash-dotted line)
signal. The grey solid line represents a smoothed power signal to
which the fit has been performed.}
\end{figure}

The strong correlation between amplitude and frequency suggests that
the signal arises from a stochastic background signal, rather than
oscillations. Stochastically excited oscillations are expected to be
excited in a particular region in the frequency spectrum, with a cut
off at the lower end instead of a continuous increase in amplitude.
To verify this, we first fitted a granulation signal to the power
spectrum of our residual signal (cf. Figure~\ref{Figure3}). The
original signal is shown in black, whereas its smoothed version is
shown as a grey solid line. The fit has been performed to the
smoothed curve assuming a granulation signal in combination with the
white noise (grey dash-dotted line line). The increasing amplitude
at lower frequencies is clearly visible, confirming the hypothesis
of a stochastic background signal. Secondly, we explore the
hypothesis of stochastically excited g-modes in more details. In the
case of stochastically excited p-modes (i.e., solar-like
oscillations), one expects a ``bump'' to occur in the power spectrum
at the location where the frequencies are excited. To check whether
this is applicable to the g-mode spectrum as well, we simulated
solar-like stochastic oscillations but with the g-mode frequencies.
As expected, a cut off at low frequencies shows up producing a clear
``bump'' in the power spectrum where the frequencies get excited.
The results of the simulation are in contradiction with what is
observed in V380\,Cyg, again implying that the observed low
amplitude signal is dominated by some stochastic ``background
noise''.

\citet{Blomme2011} investigated three O-type stars, HD\,46223,
HD\,46150, and HD\,46966, based on high-quality light curves
gathered with the CoRoT \citep{Auvergne2009} space mission. The
three studied objects are located in the HR diagram just at the
boundary of the instability strip due to the $\kappa$-mechanism.
However, the amplitude spectra of all three stars were found to be
unlike any of known types of pulsating stars. That is, the spectra
show a clear trend of increasing amplitude as the frequency
decreases; the individual frequencies are found to vary both in
appearance and in time, suggesting a stochastic nature of the
signal. The authors concluded that most of the power detected in the
periodograms of these three stars can be ascribed to ``red noise'',
which has nothing to do with the instrumental noise but rather
points to (predominantly) stochastic behaviour. This red noise is
assumed to be intrinsic to the star and \citet{Blomme2011} discussed
three possibilities for its physical cause: sub-surface convection,
granulation, and inhomogeneities in the stellar wind. These authors
stressed that all three options are highly speculative, although
previous studies already showed a link between convection and red
noise \citep{Schwarzschild1975}, reported about observational
evidence of clumping in stellar winds of early-type stars
\citep{Puls2008}, as well as the existence of a sub-surface
convective layer that manifests itself as a large microturbulent
velocity field, the generation of magnetic fields, and line-profile
variations \citep{Cantiello2009}.

\begin{table*} \centering
\tabcolsep 3.5mm\caption{\label{Table: orbital_elements} Comparison
of the orbital elements of the V380\,Cyg system derived by us with
the results of previous studies. $T_{\rm 0}$ (24\,00000+) stands for
the time of periastron passage, $e$ for the orbital eccentricity,
$\omega$ for the longitude of the periastron, $q$ for the mass
ratio, and $K_{\rm A}$ and $K_{\rm B}$ for the RV-semiamplitude of
the star A and B, respectively. Our errors are represented by
standard deviations of the mean and thus do not resemble real error
bars.}
\begin{tabular}{llccccc}
\hline
\multirow{2}{*}{Parameter}  & \multirow{2}{*}{Unit} &     Hill \& Batten       & Lyubimkov     &  Guinan         & Pavlovski      &  Present  \\
                            &                       &        (1984)            & et al.\ (1996) &  et al.\ (2000) &   et al.\ (2009) &     work    \\
\hline
$T_{\rm 0}$ & JD             & 37455.10$\pm$0.11  & 37454.974$\pm$0.054 &  49495.70$\pm$0.09 & 54615.18$\pm$0.14 &    54602.888$\pm$0.007  \\
 $e$        &                & 0.22$\pm$0.01      & 0.2183$\pm$0.0051   & 0.22               & 0.206$\pm$0.008 &     0.2261$\pm$0.0004    \\
 $\omega$ & degree             & 127.6$\pm$2.8      & 122.2$\pm$1.8       & 132.7$\pm$0.3      & 134.2$\pm$1.1  &  141.5$\pm$0.2     \\
 $K_{\rm A}$ & \kms & 92.8$\pm$2.6       & 93.95               & 95.6$\pm$0.5       & 95.1$\pm$0.3  &  93.54$\pm$0.07  \\
 $K_{\rm B}$ & \kms & 160.7$\pm$2.8      & 155.3               &147.9$\pm$1.2       & 160.5$\pm$1.2 & 152.71$\pm$0.22  \\
 $q$ &                        &  0.577$\pm$0.019             &  0.605              & 0.646$\pm$0.004  &  0.592$\pm$0.005  & 0.613$\pm$0.001 \\
\hline
\end{tabular}
\end{table*}

\citet{Neiner2012} presented a study of the CoRoT Be star HD\,51452.
The authors reported the detection of rather high amplitude
(0.5--1.0 mmag) oscillations in the g-mode regime for this B0 star,
whereas no (significant) power in the p-mode regime is present. They
found that the star has close to solar chemical composition and is
too hot for g-modes to be excited by the $\kappa$-mechanism.
\citet{Neiner2012} proposed a stochastic excitation mechanism in the
gravito-inertial regime as the most realistic explanation of the
low-frequency oscillations detected in CoRoT photometry of
HD\,51452, where the modes are assumed to be excited either in the
convective core or in a thin convective sub-surface layer. Though
the shape of the amplitude spectrum of HD\,51452 is different from
those reported by \citet{Blomme2011} for three O-type stars in the
CoRoT field, both studies concluded that theoretically predicted
modes could not be detected in their objects, and that the observed
frequencies are too much off from the theoretical ones.

\citet{Shiode2013} investigated the excitation and propagation of
stochastic internal gravity waves as well as their detectability in
terms of flux and RV variations at the surfaces of massive stars.
The authors found that the convective core of high-mass,
main-sequence stars might be responsible for the excitation of these
modes which cause intrinsic photometric variability at frequencies
of 0.4--0.8~\cd\ (5--10~\mhz) with surface amplitudes of some tens
of micromagnitudes. As the stars evolve along the main-sequence,
oscillations at frequencies down to 0.08~\cd\ ($\sim$1~\mhz) become
detectable and the surface flux variability reaches the level of
some hundred micromagnitudes. These theoretical predictions are in
agreement with the results of \citet{Blomme2011}, who detected power
at the same frequencies and amplitudes in amplitude spectra of three
O-type stars. However, the theory is found to disagree with the
observations regarding the mode lifetimes: while \citet{Blomme2011}
found mode lifetimes to be of the order of days to hours, the
theoretical predictions of \citet{Shiode2013} suggested lifetimes of
years to Myrs. \citet{Rogers2013} also found that internal gravity
waves can effectively carry angular momentum within a star and might
cause spin up or down of the outer layers of the star and might be
responsible for the Be phenomenon observed in some B-type stars.

The signal we observe in the primary component of V380\,Cyg is very
similar to that observed by \citet{Blomme2011} in their three O-type
stars: it is clearly not instrumental, it is variable both in
appearance and amplitude on short time-scales \citep[see Figure~3
in][]{Tkachenko2012}, and it has low amplitude at the level of some
hundred micromagnitudes. These characteristics of the observed
signal are not in contradiction with the theoretical predictions by
\citet{Shiode2013} and \citet{Rogers2013} for stochastically excited
gravity waves. We will show that the primary of V380\,Cyg exhibits a
high microturbulent velocity. Both \citet{Cantiello2009} and
\citet{Shiode2013} showed that thin convective sub-surface layers
present in some massive stars can be another important source of
g-modes near the stellar surface. The authors found that these modes
are characteristic of degree $l>$30, which implies that they cannot
produce large surface brightness variations, but still could be
responsible for large microturbulent fields observed in massive
stars. The high microturbulent velocity observed in the primary of
V380\,Cyg supports this theory. On the other hand, the large
microturbulent velocity field can only be considered as an indirect
confirmation of the theory, while such high degree non-radial modes
cannot be directly observed because of their very low amplitudes in
the disk integrated light.

\section{Spectroscopic analysis}

In this Section, we present a detailed spectroscopic analysis of the
V380\,Cyg system based on the high-quality spectroscopic data
described in Section~2. The analysis includes determination of the
orbital parameters and the disentangled spectra of both components
by means of the spectral disentangling technique, estimation of the
atmospheric parameters and chemical composition for both stars as
well as further comparison with the evolutionary models, and the
frequency analysis and the Doppler Imaging analysis of the selected
spectral lines in the spectrum of the primary component.

\subsection{Spectral disentangling}

We use the spectral disentangling ({\sc spd}) method
\citep{Simon1994} to determine the orbital elements for the V380~Cyg
system. The method enables simultaneous separation of the intrinsic
individual spectra of the components and determination of the
orbital parameters in a self-consistent way from time-series of the
observed composite spectra of a binary system. In this sense, the
measurements of the RVs of the components from usually complex
blends of spectral lines are by-passed, and a set of the orbital
elements is optimised instead.

{\sc spd} has been performed in Fourier space as formulated by
\citet{Hadrava1995} and implemented in the {\sc FDBinary} code
\citep{Ilijic2004}. Such implementation is less demanding on the
memory storage, and on the CPU time than the original method
introduced by \citet{Simon1994}. The Discrete Fourier Transform
(DFT) is used in the {\sc FDBinary} code which enables more freedom
in the selection of the spectral regions as well as an arbitrary
number of selected pixels which allows better preservation of the
original spectral resolution of the input spectra.

Spectral lines of both stars get distorted in the course of the
corresponding eclipses due to the Rossiter-McLaughlin effect
\citep{Rossiter1924,McLaughlin1924}. Therefore, the principal
condition of spectral disentangling stating that no variation other
than that due to the orbital motion should be present in the spectra
is violated at these phases. Hence, only out-of-eclipse spectra of
V380\,Cyg have been selected by us for the {\sc spd} analysis.
However, without any significant change in the fractional light
contribution of the components in the course of the orbital cycle,
the zero-order Fourier mode becomes singular. This leads to an
ambiguity in the obtained disentangled spectra - external
information on the light ratio is needed for their proper
renormalization
\citep[cf.,][]{Pavlovski2005,Pavlovski2010,Pavlovski&Southworth2009}.
Moreover, undulations in the disentangled spectra might appear but
this unwanted disturbance could be handled with appropriate phase
coverage of the input spectra \citep{Hensberge2008}.

\begin{figure*}
\includegraphics[width=175mm]{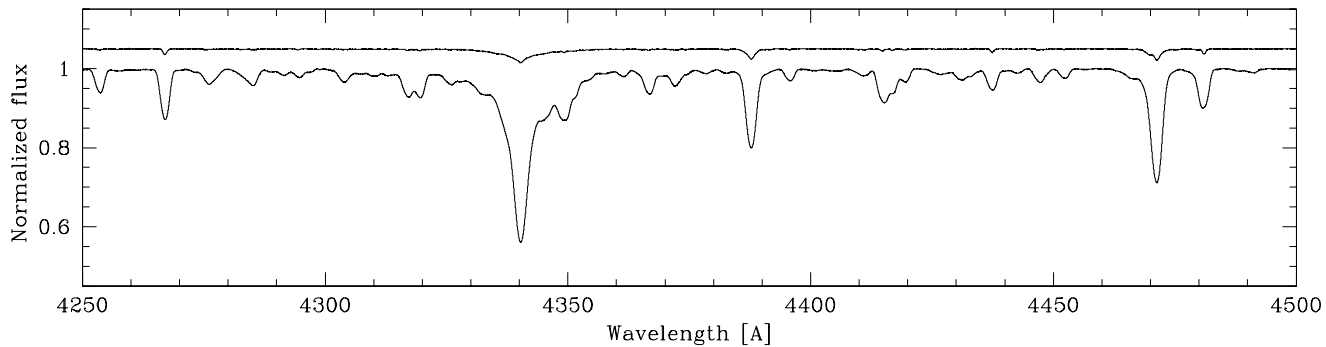}
\caption{\label{Figure: Decomposed_spectra} Disentangled spectra of
both components of the binary system V380\,Cyg in the region of
H$\gamma$, C~{\small II}~4267~{\AA}, and He~{\small I}~4388 and
4471~{\AA} spectral lines. The spectrum of the secondary was shifted
up in the vertical direction for clarity.}
\end{figure*}

Out of 452 spectra, 237 measurements were taken at the
out-of-eclipse phases. The optimisation in the {\sc FDBinary} code
is performed by means of the simplex algorithm. In order to avoid
possible local minima in the solution, we ran a very large number of
independent runs which makes {\sc spd} quite demanding on CPU time
even with the reduced (to 237) number of input spectra. From the
previous studies (G2000, P2009, Paper~I), it is known that the
effective temperatures of the components are similar ($\sim
22\,000$~K), and are in the range of the maximum intensity of
He~{\small I} lines. This is a fortunate situation for the quality
of the determination of the orbital parameters in the case of a
large light ratio between the components. Mainly due to their large
intrinsic width, Balmer lines are not suitable for the accurate
determination of the orbital elements. Thus, the orbital elements
were derived by focusing on several wavelength intervals centred at
the prominent He~{\small I} lines at $\lambda\lambda$ 4026, 4388,
4471, 4713, 4920, and 6678~${\AA}$; the mean orbital elements are
listed in the last column of Table~\ref{Table: orbital_elements}.
The errors quoted are the standard deviations of the mean. These are
not realistic (systematic) errors but rather illustrate an excellent
consistency between the orbital elements derived from different
spectral regions.

A portion of the components' disentangled spectra centred at
H$\gamma$ and including two He~{\small I} and one C~{\small II}
lines is illustrated in Figure~\ref{Figure: Decomposed_spectra}. The
faintness of the secondary (upper spectrum) is obvious despite its
similar effective temperature to the primary's. We used the orbital
solution from Table~\ref{Table: orbital_elements} (last column) to
do the spectral disentangling in a wide wavelength range that,
besides helium and metal lines, also included three Balmer lines
(H$\delta$, H$\gamma$, and H$\beta$). We did not consider the red
part of the optical spectrum beyond 580~nm for the {\sc spd} to
prevent the influence of the telluric spectrum on the resulting
disentangled spectra. Only those few regions the least influenced by
the telluric contributions and centred at some isolated and
prominent metal (or helium) lines were subjected to spectral
disentangling in that part of the spectrum (e.g., a couple of strong
lines of C~{\small II} at 6578 and 6582~{\AA}, He~{\small I} line at
6678~{\AA}, etc.).

Table~\ref{Table: orbital_elements} compares our orbital solution
with the most recent findings from the literature. A comparison
between the different solutions is not straightforward, however,
mainly because the corresponding studies were based on spectra of
different quality and quantity. The largest deviations are found for
the RV-semiamplitude of the secondary component, $K_{\rm B}$, and
thus for the mass ratio, $q$. The main reason for this is the small
spectral contribution of the main-sequence companion to the
composite spectra of the binary, given its faintness compared to the
evolved primary component. Indeed, one of the most recent studies by
P2009 involving high-resolution spectroscopy was based on the same
method of {\sc spd} but provided a significantly higher value of
$K_{\rm B}$ (and thus a lower mass ratio) compared to our findings.
We suppose the main reason for this discrepancy is a lower S/N of
and a poor orbital phase coverage provided by the spectra analysed
in P2009. Herewith, with about 240 spectra used for the analysis, we
have almost perfect phase coverage, except for the omitted spectra
at the phases of both eclipses.

\begin{figure*}
\includegraphics[width=55mm]{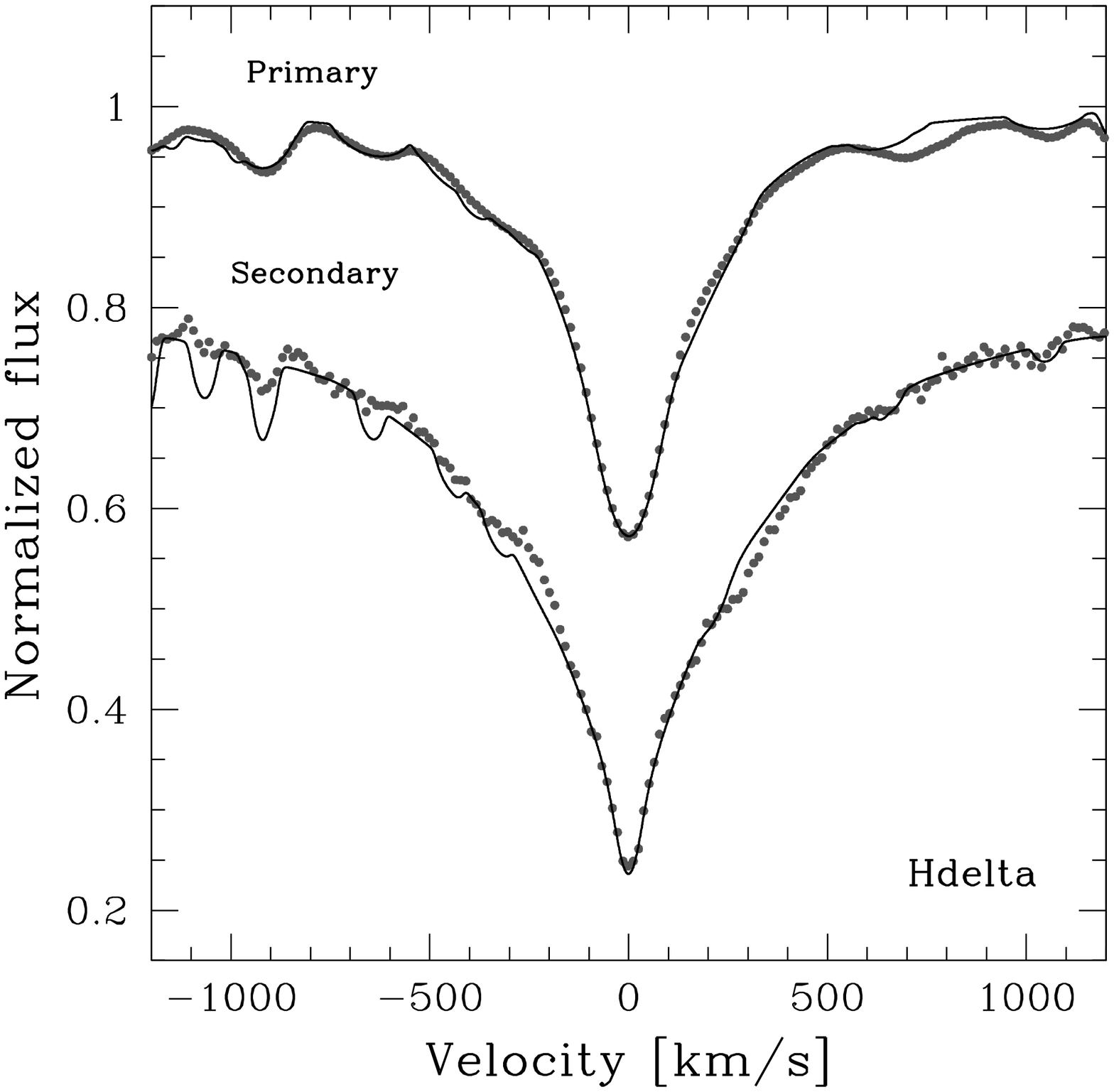}\hspace{5mm}  
\includegraphics[width=55mm]{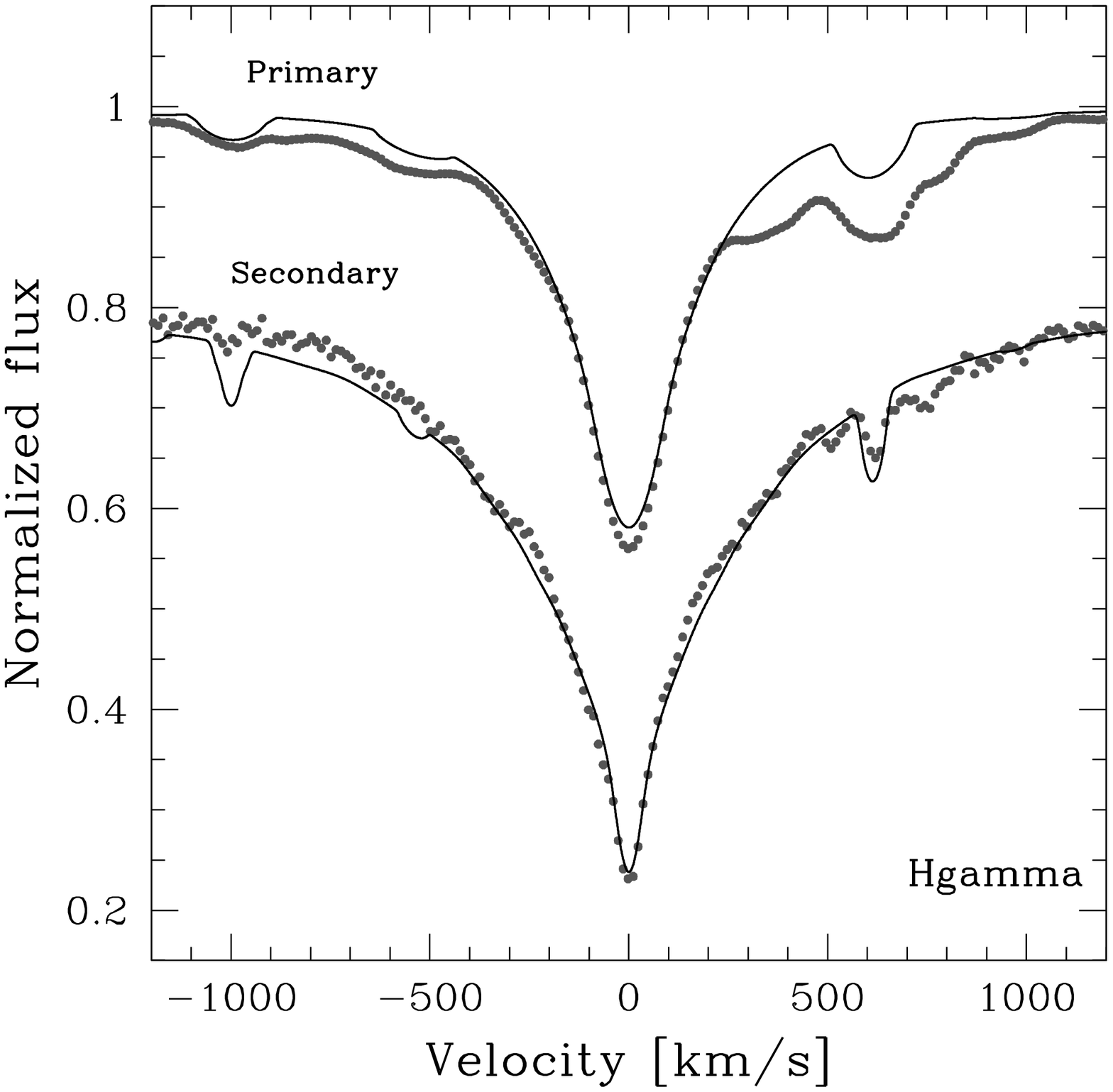}\hspace{5mm}  
\includegraphics[width=55mm]{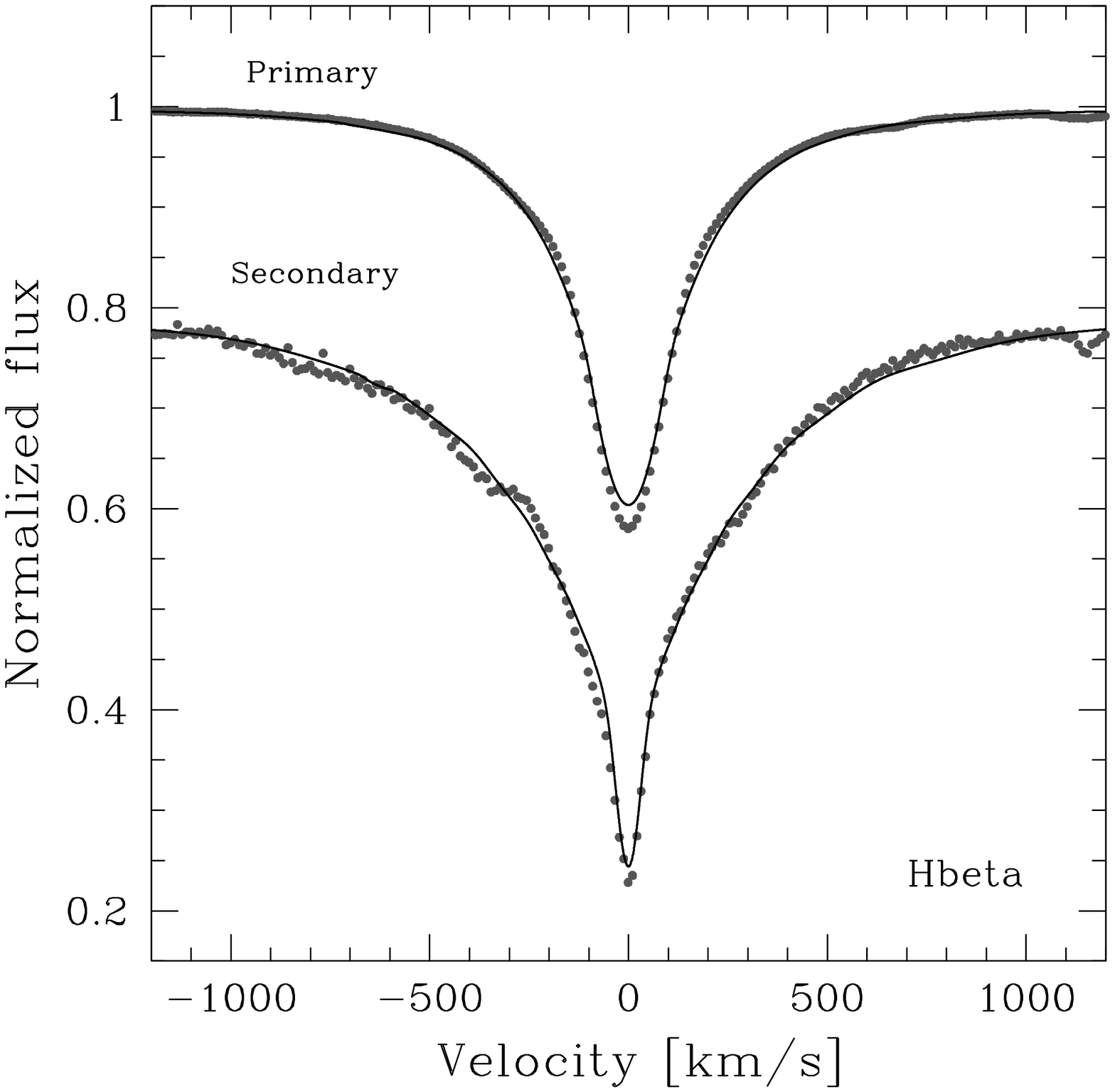}              
\caption{\label{Figure: Balmer_lines_fit} The best fit synthetic
spectra (lines) in comparison to the renormalized disentangled
spectra (dots) of the primary (upper) and the secondary component
(lower). From the left to right panel H$\delta$, H$\gamma$, and
H$\beta$ profiles are shown.}
\end{figure*}

\subsection{Spectrum analysis of both binary components}

As mentioned above, the {\sc spd} method, along with the orbital
solution, provides disentangled spectra of the individual stellar
components of a binary system. The spectra can be analysed by
methods suitable for single stars to determine the effective
temperatures, surface gravities and chemical composition of the two
stars to a high precision
\citep{Hensberge2000,Pavlovski2005,Fremat2005,Clausen2008,Hareter2008,
Pavlovski&Southworth2009,Tkachenko2009}. In particular, the
effective temperature determined from hydrogen line profiles, and/or
strength of the temperature sensitive lines, superseded in accuracy
and precision those derived from e.g., broad-band or Str\"{o}mgren
photometry. The same holds for the metallicity derived from detailed
a abundance study of stellar spectra rather than photometric colour
indices.

Determination of the effective temperature from hydrogen lines of
the Balmer series in OB stars suffers from degeneracy in \te\ and
\logg. This degeneracy is a major limitation in setting-up of the
appropriate model atmosphere for a detailed abundance study.
Thankfully, complementary analysis of the light and RV curves
(variations) in detached eclipsing binaries enables the
determination of the components' masses and radii with high
accuracy. Knowledge of these two fundamental characteristics of the
star ensures that the surface gravity is also known to a high
precision (~0.01~dex or better). Hence, the above mentioned
degeneracy in \te\ and \logg\ can be safely removed for this type of
stars.

In the present analysis, we renormalize the disentangled spectra
using the light ratio ($l_{\rm B}/l_{\rm A} = 0.062\pm0.001$)
derived in the course of the light curve analysis (cf. Section 3.1).
The effective temperatures of the components are then determined by
means of the optimal fitting of the renormalized spectra of the
individual stellar components to a grid of theoretical spectra.
Since we are dealing with B2 spectral type stars, the synthetic
spectra were computed assuming non-LTE line formation (where ``LTE''
stands for the {\it local thermodynamical equilibrium}). However,
the whole analysis is based on the so-called {\it hybrid approach}
that assumes LTE-based atmosphere models computed with the {\sc
atlas9} code \citep{Kurucz1993} and non-LTE spectral synthesis with
detailed statistical balance by means of the {\sc detail}
\citep{Butler1984} and {\sc surface} \citep{Giddings1981} codes.
Justification of such approach has been discussed in detail in
\citet{Nieva2007}. The grid of theoretical spectra ranges from
18\,000 to 25\,000~K with a step width of 1\,000~K in effective
temperature and from 3.0 to 4.5~dex with a step size of 0.1~dex in
surface gravity. Details on input atomic data are given in
\citet{Nieva2012}.

The optimal fitting was performed with the genetic algorithm as
implemented in the {\sc pikaia} subroutine after
\citet{Charbonneau1995}. Besides the effective temperature for each
binary component, we also optimised for the projected rotational
velocities, the velocity offset between the disentangled spectra and
the rest-frame wavelengths of the theoretical spectra, and the
continuum correction (vertical shift in flux). In practice, echelle
blaze orders that contain broad Balmer lines are difficult to
normalize and merge. In turn, the corresponding spectral regions are
the main source of difficulties in the {\sc spd}. In particular, in
Fourier-based disentangling, wrong continuum placement usually shows
up as undulations in the resulting disentangled spectra
\citep{Hensberge2008}. Moreover, due to the fact that we assume the
light ratio between the two components to be time-independent, the
disentangled spectra are not properly normalized and an additive
correction value should be applied to return them to the level of
unity.

Table~\ref{Table: atmospheric_parameters} lists the derived
atmospheric parameters along with some fundamental quantities
derived from the light curve fitting (cf. Section 3.1). The
parameters derived spectroscopically are marked with an asterisk
(``*''). Effective temperatures were determined from fitting the
Balmer lines whereas \vsini\ and $v_{\rm turb}$ were estimated from
He~{\small I} and metal lines. Similar to the uncertainties on the
orbital elements (cf. Section~4.1), the quoted errors are the
standard deviations of the mean. As such, small error bars for the
effective temperature of the primary in Table~\ref{Table:
atmospheric_parameters} stand for an excellent consistency between
the values derived from different Balmer lines. Figure~\ref{Figure:
Balmer_lines_fit} compares the best fit synthetic spectra for both
components of V380\,Cyg (upper and lower profiles for the primary
and secondary, respectively) with the corresponding disentangled
spectra, renormalized to the individual continua, in the regions
centred at three Balmer lines: H$\delta$ (left), H$\gamma$ (middle),
and H$\beta$ (right). The difference in S/N between the renormalized
spectra of the components is obvious and is due to small
contribution of the secondary to the total light of the binary
system. The derived effective temperature of the primary compares
well to the values reported by G2000 and P2009 (cf.
Table~\ref{Table1}). Moreover, both authors report about the high
value of microturbulent velocity $v_{\rm turb,A}$ from 12 to 14
\kms\ (G2000 and P2009, respectively), though disagree on the
metallicity value: $-0.44\pm$0.07~dex derived by G2000 to be
compared to $-0.02\pm$0.05~dex found by P2009. We confirm the high
value of the microturbulent velocity: $v_{\rm turb,A} =
15\pm$1~\kms\ from our spectra. Our abundance analysis suggests a
similar chemical composition for both binary components (see below);
the magnesium abundance of the primary is consistent with the
metallicity index reported by P2009.

For the secondary component of V380\,Cyg, our spectroscopic and
photometric data suggest significantly different values for the
effective temperature. While the optimal fitting of the disentangled
spectrum of the star results in $T_{\rm eff, B} = 21\,800\pm400$~K,
a value only slightly above the one found for the primary component,
the light curve modelling delivers the value of $T_{\rm eff,B} =
23\,840\pm500$~K (assuming fixed to the spectroscopic value
temperature of the primary, cf. Table~\ref{Table:
atmospheric_parameters}). Since the disentangled spectra are
renormalized using the photometric value of the light ratio between
the two stars, we are somewhat in a ``dead cycle'': the high
photometric effective temperature is inconsistent with the spectral
characteristics of the star. As will be shown in Section~4.5,
adopting higher effective temperature for the secondary leads to
large discrepancy of $\sim$2~M$_{\odot}$ between the mass deduced
from binary dynamics and the one obtained from evolutionary models.
As a compromise between the two temperatures, we finally adopt the
effective temperature of the secondary to be $T_{\rm eff,B} =
22\,700\pm1200$~K, i.e. the mean of the spectroscopic and
photometric values. Besides \te\ and \vsini, we also derived the
micorturbulent velocity for the secondary to be  $v_{\rm turb,B} =
1\pm$1~\kms. This parameter is determined for the secondary for the
first time; the low value is not unexpected for a B-type star at
early phases of the evolution on the main-sequence
\citep{Lefever2010,Nieva2012,Lyubimkov2013}.

\begin{table} \tabcolsep4.0mm \centering \caption{\label{Table: atmospheric_parameters}
Atmospheric parameters of both components of V380\,Cyg along with
some fundamental characteristics derived from the light curve
fitting. \Veq\ and \Vsync\ are the observed equatorial and
calculated synchronous rotational velocities, respectively. Error
bars are 1-$\sigma$ level, with $\sigma$ the standard deviation of
the mean.}
\begin{tabular}{l l r@{\,$\pm$\,}l r@{\,$\pm$\,}l} \hline
 Parameter & Unit                                    &      \mc{Star A}      &      \mc{Star B}      \\
\hline
\logg$^*$ & cm\,s$^{-2}$                      & 3.104      & 0.006     & 4.120     & 0.011     \\
\te & K           & 21\,700    & 300       & 22\,700   & 1200       \\
$v_{\rm turb}$ & \kms & 15 & 1 & 1 & 1\\
\vsini & \kms                             & 98         & 2         & 38 & 2\\
\Veq & \kms                        & 99.3         & 1.0       & 38.5     & 3.0        \\
\Vsync & \kms                      & 63.7       & 0.5       & 15.6     & 0.3      \\[2pt]
\hline \multicolumn{6}{l}{$^*$ from the light curve solution}
\end{tabular}
\end{table}

\begin{figure}
\includegraphics[width=40mm]{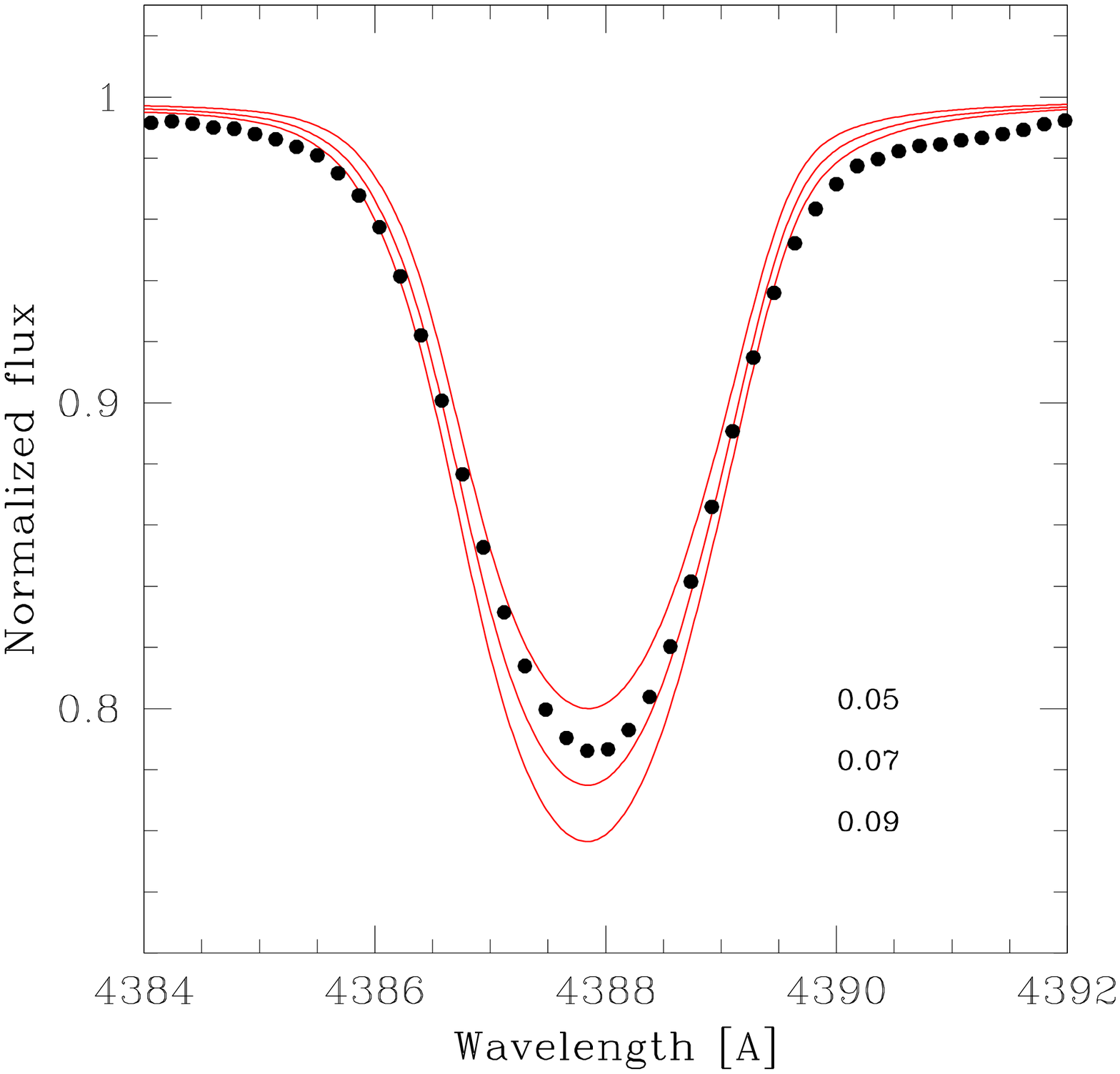}\hspace{3mm}
\includegraphics[width=40mm]{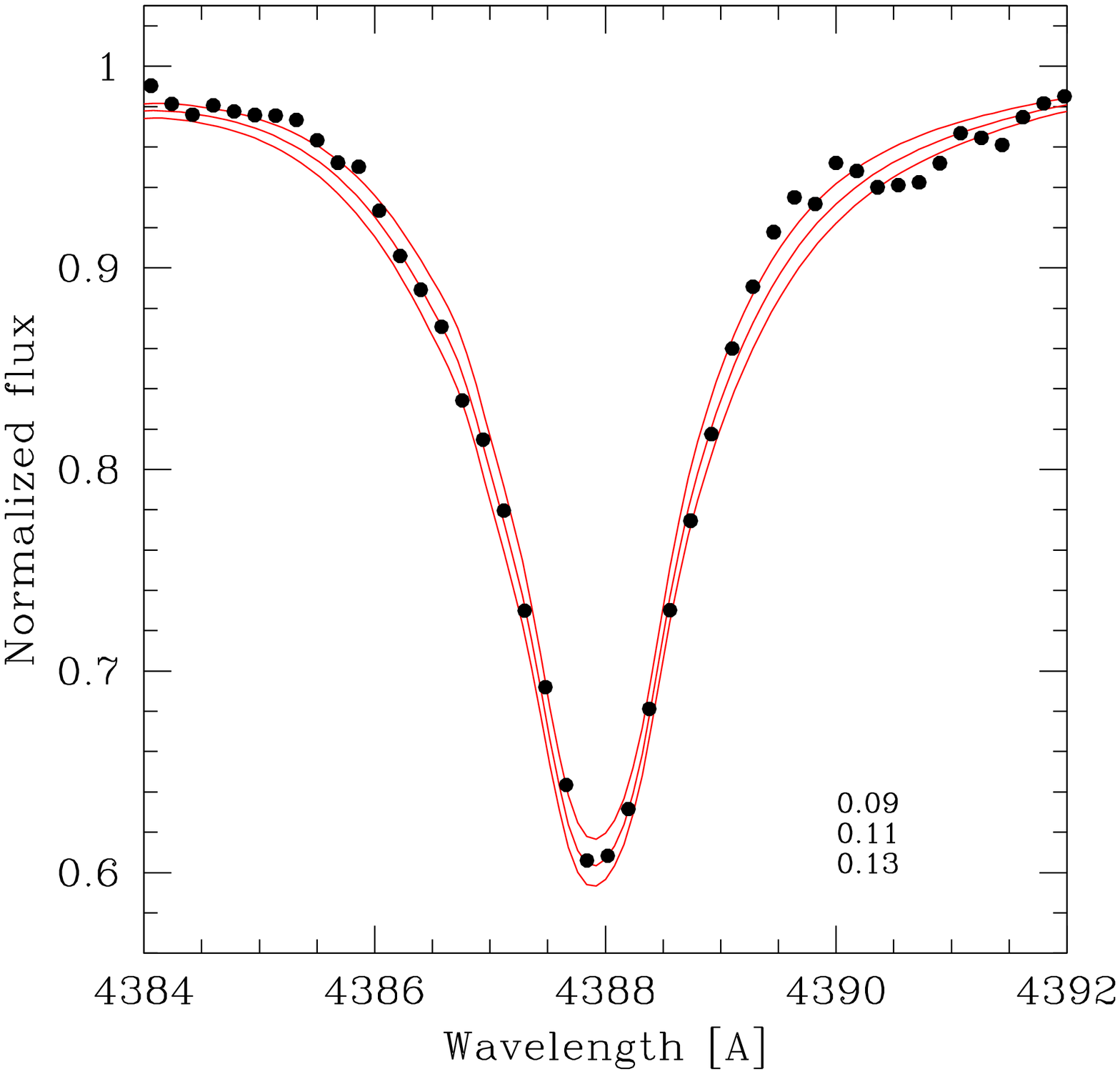}
\includegraphics[width=40mm]{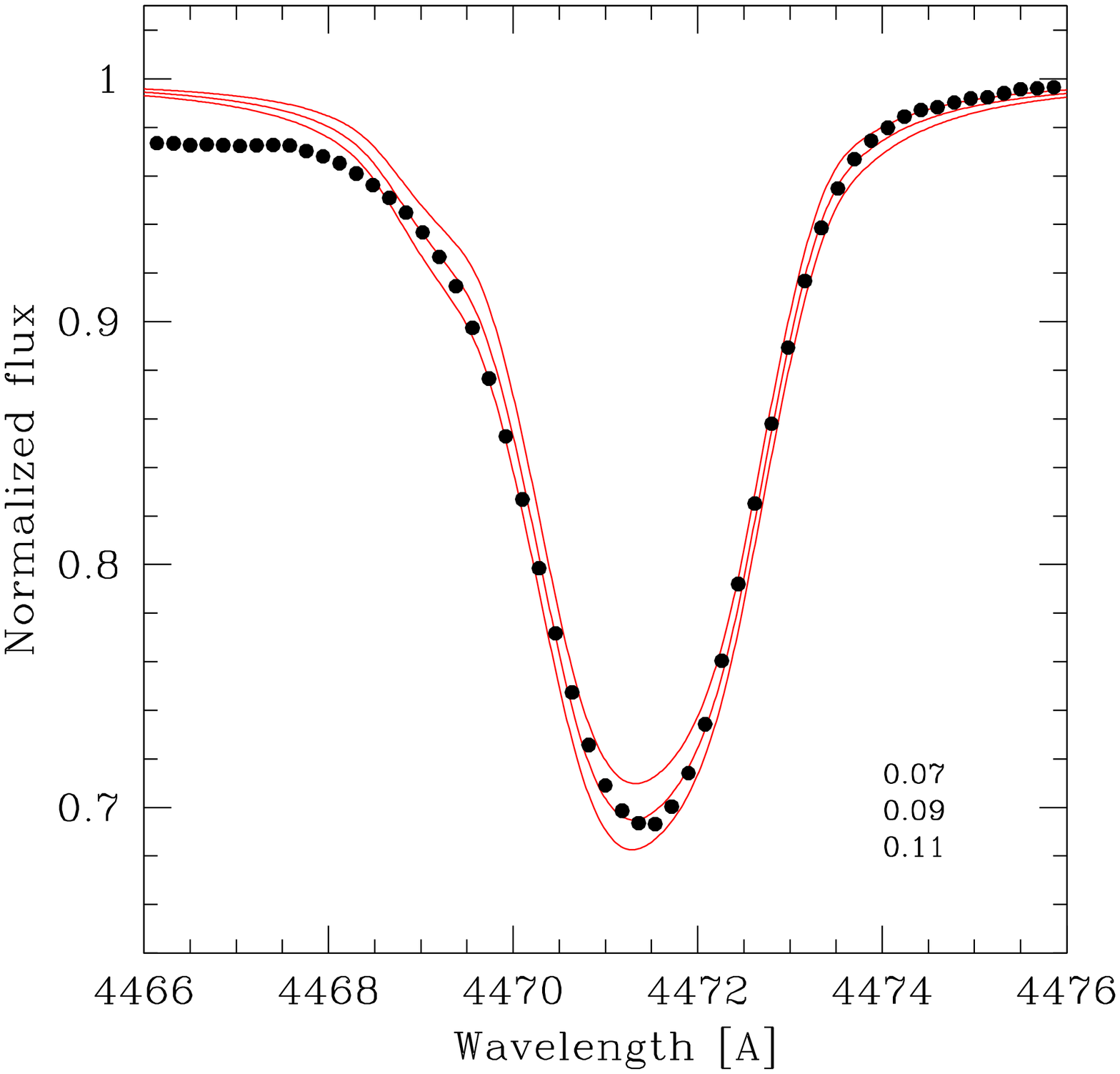} \hspace{2mm}
\includegraphics[width=40mm]{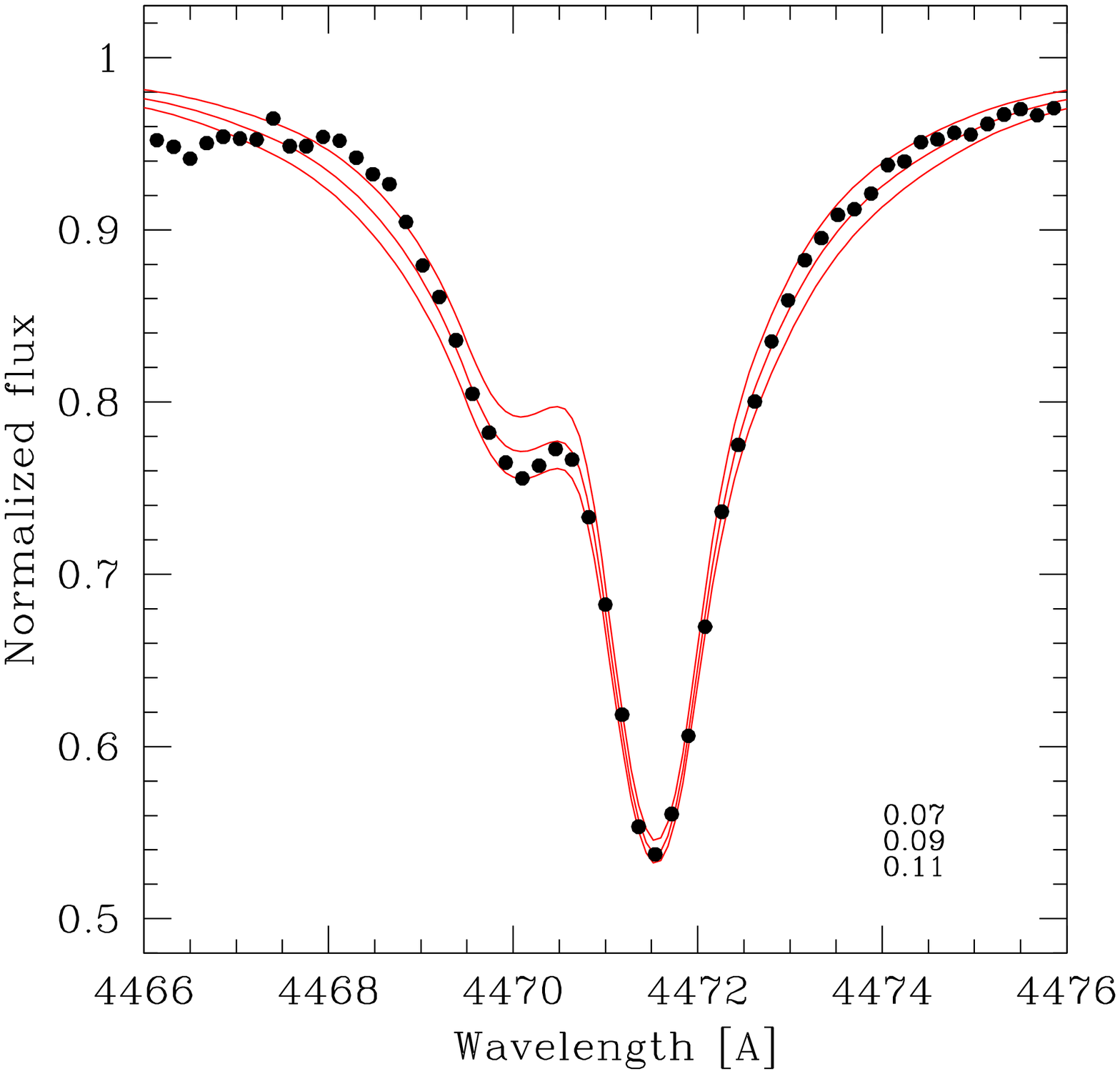}
\includegraphics[width=40mm]{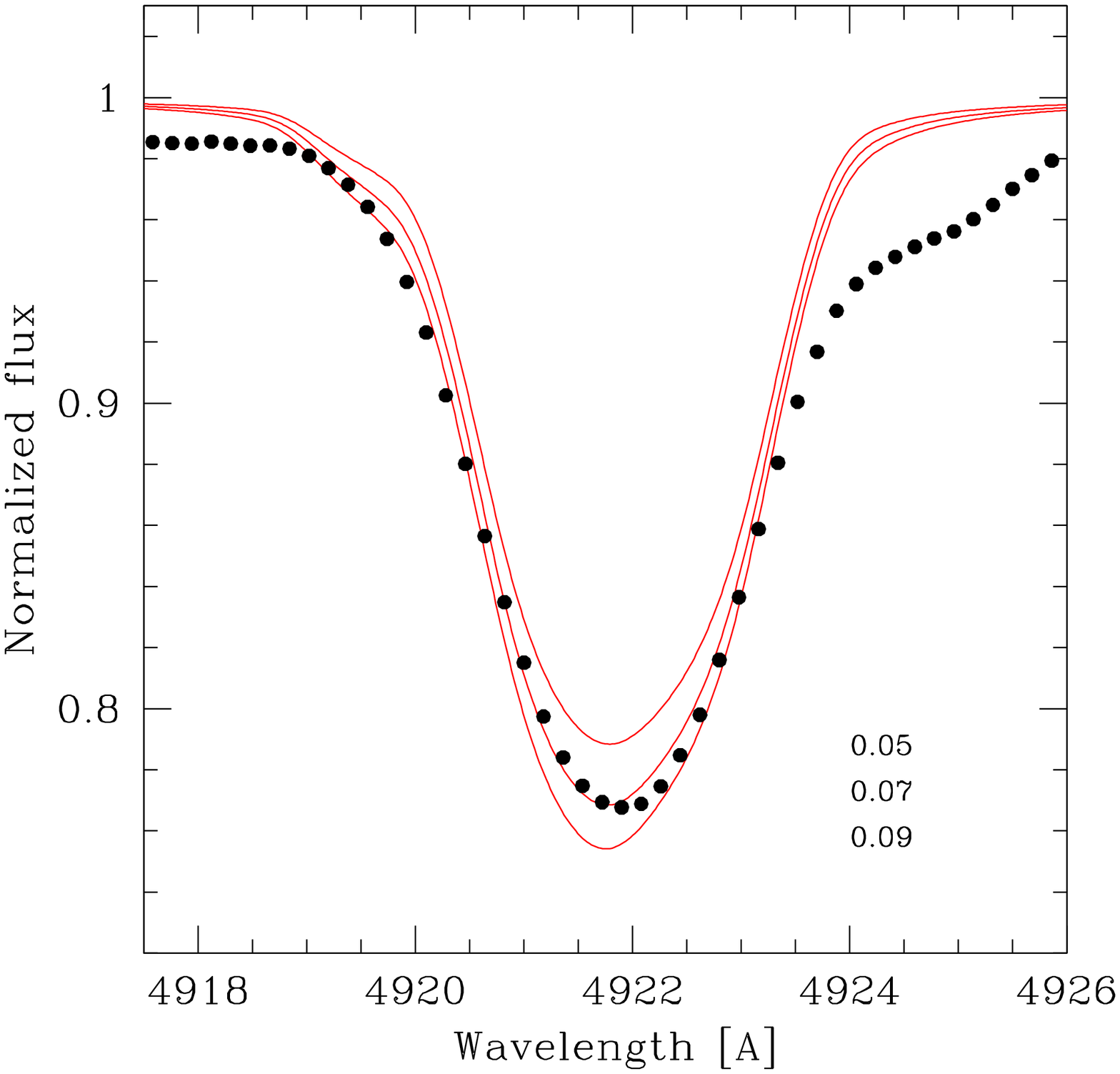} \hspace{3mm}
\includegraphics[width=40mm]{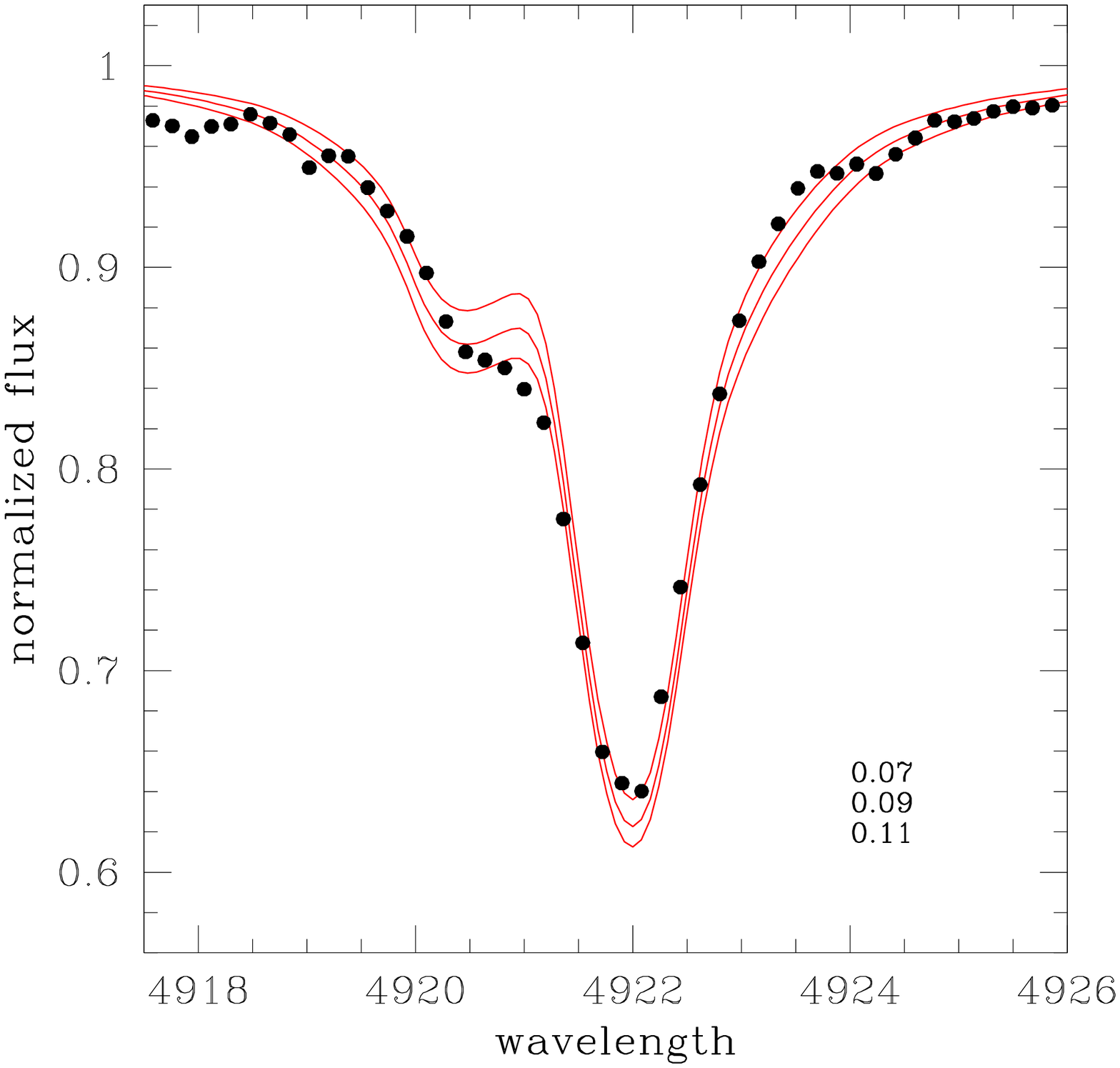}
\caption{\label{Figure: Helium_lines_fit} Theoretical spectra
(lines) of He~{\small I} lines at 4387.9, 4471.5 and 4921.9 {\AA}
(from top to bottom)  compared to the renormalized disentangled
spectra (dots) of the primary (left panels) and the secondary
components (right panels). Theoretical spectra are calculated for
the stellar parameters listed in Table~\ref{Table:
atmospheric_parameters}; the assumed different helium abundances are
indicated in the bottom right corner of each panel. The colour
version of the figure is available online only.}
\end{figure}

Table~\ref{Table: helium_abundances} lists helium abundances for
both stellar components of V380\,Cyg. The abundances were measured
from ten individual He~{\small I} lines in the corresponding
disentangled spectra. Within the quoted (1-$\sigma$) errors, helium
abundances of the two binary components agree with each other as
well as with the value derived by \citet{Nieva2012} from a sample of
OB stars, but are slightly higher than the solar helium abundance
found by \citep{Asplund2009}. The quality of the fits for a few
selected helium lines in the spectra of both components are shown in
Figure~\ref{Figure: Helium_lines_fit}. Our results agree with the
findings by P2009 in the sense that line-to-line scatter in helium
abundance is higher than anticipated. On the other hand, these
authors report slightly higher mean helium abundance of
0.094$\pm$0.031~dex for the primary component derived from the same
set of spectral lines, with the two He~{\small I} lines at 5875.7
and 6678.1 {\AA} showing the highest abundances. A small difference
between our and P2009 values can be attributed to slightly different
microturbulent velocities derived for the primary component as well
as to the difference in the light ratio that was used to renormalize
the disentangled spectrum of the star.

Helium abundances of both binary components were also determined by
\citet{Lyubimkov1996} based on the equivalent widths measured
directly from the composite spectra of the binary. Given a small
contribution of the secondary component of $\sim$6\% to the total
light, this way of measuring abundances for this star is associated
with large uncertainties. The authors reported a strong helium
overabundance of $\sim$0.16$\pm$0.04~dex for the primary component,
with rather large line-to-line scatter. For the secondary, they
reported a value of 0.074$\pm$0.009~dex which is lower than both the
solar value found by \citet{Asplund2009} and the one derived for a
sample of OB stars by \citet{Nieva2012}. \citet{Lyubimkov1996}
interpreted their distinctly high helium abundance for the primary
component as evidence for a strong internal mixing by turbulent
diffusion, which brings CNO processed material from the core to the
surface already during the main-sequence phase. The helium
abundances of the binary components derived in our study are in
contradiction with the findings and conclusions by
\citet{Lyubimkov1996}. Similar to the components of V380\,Cyg, no
excess helium abundance was found in several other high-mass stars,
members of binary systems \citep[e.g.,][]{Pavlovski2005,
Pavlovski2009, Pavlovski&Southworth2009,Mayer2013}.

\begin{figure*}
\includegraphics[width=84mm]{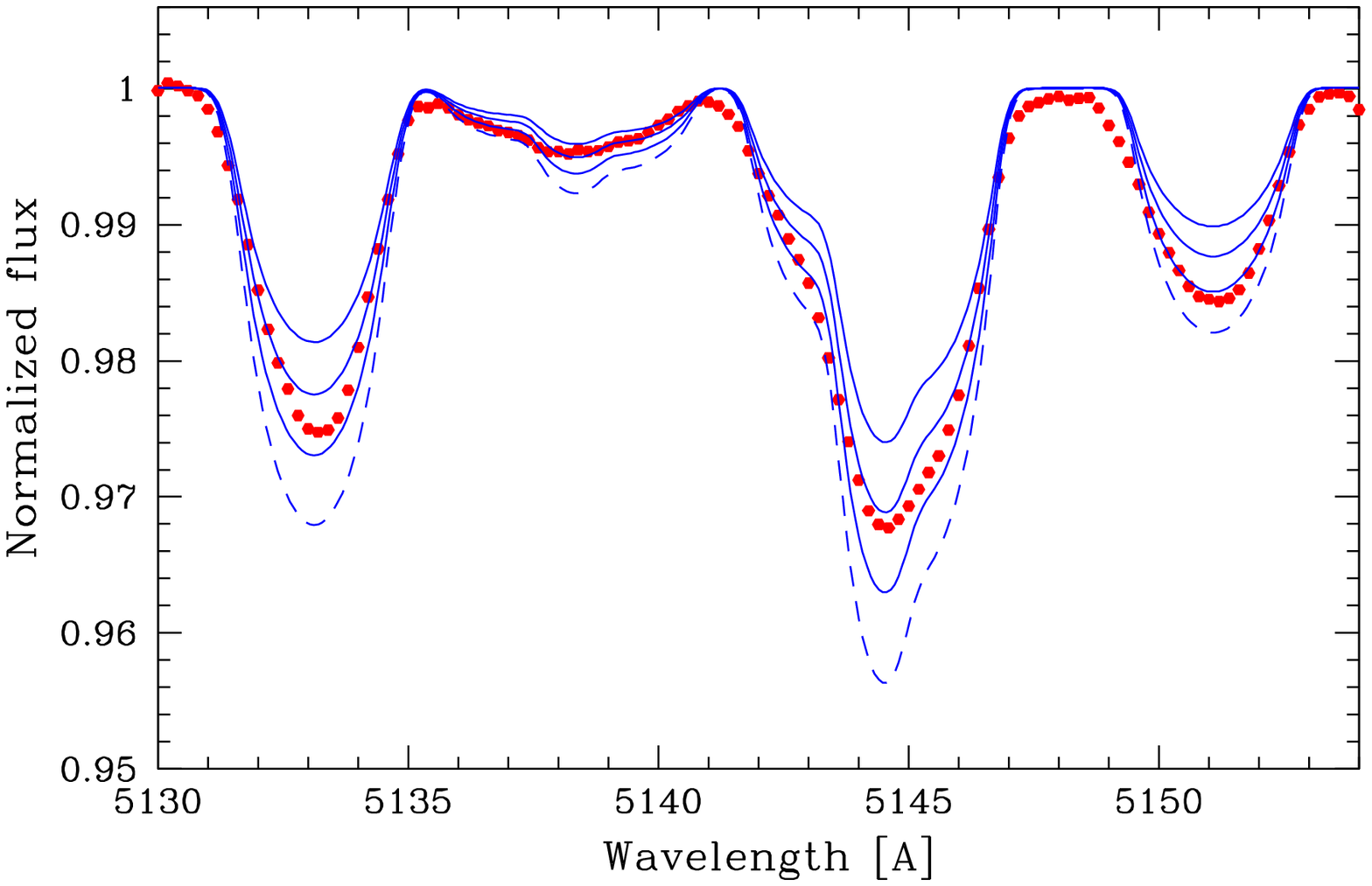}\hspace{5mm}
\includegraphics[width=84mm]{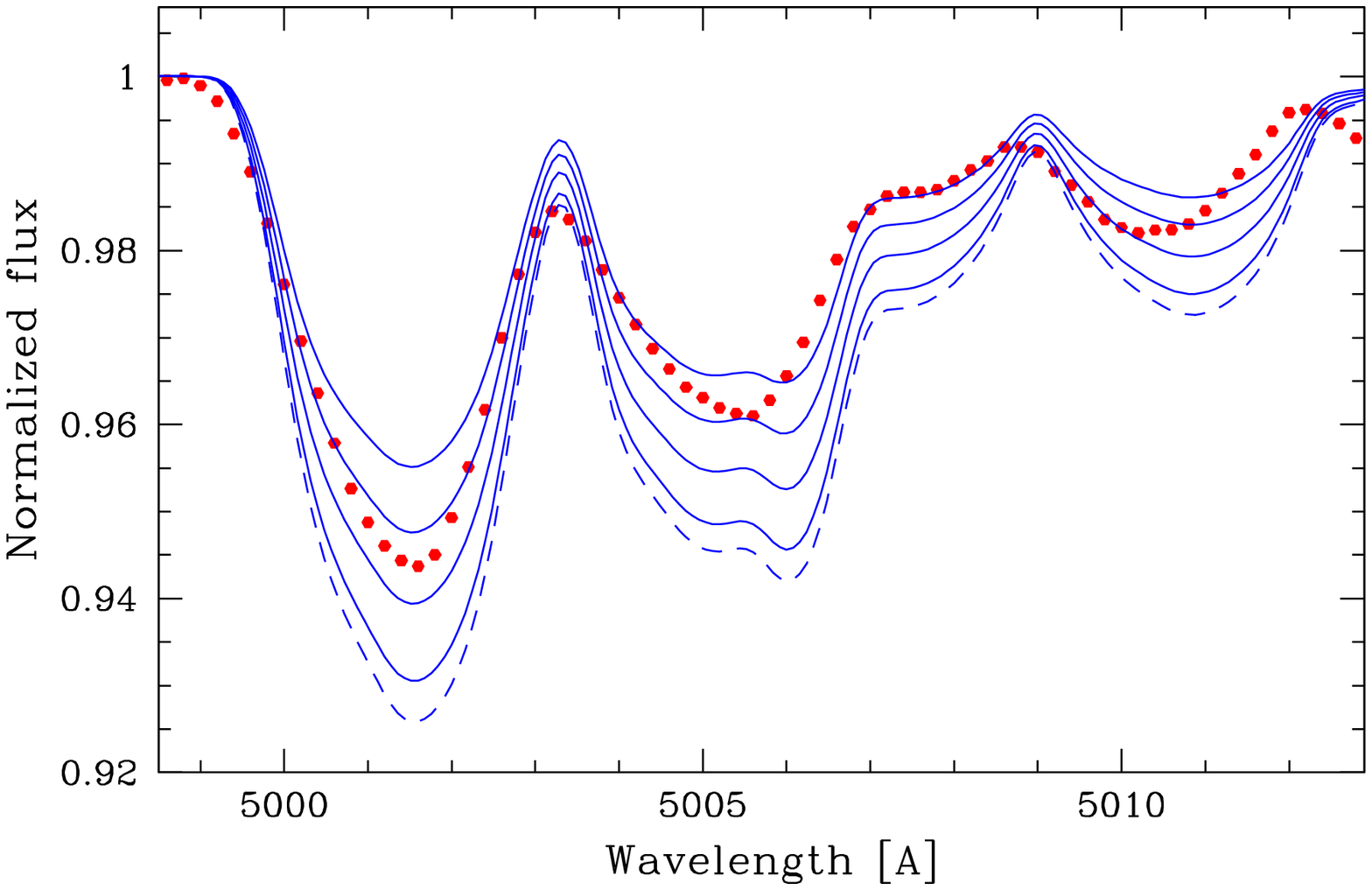}\vspace{5mm}
\includegraphics[width=84mm]{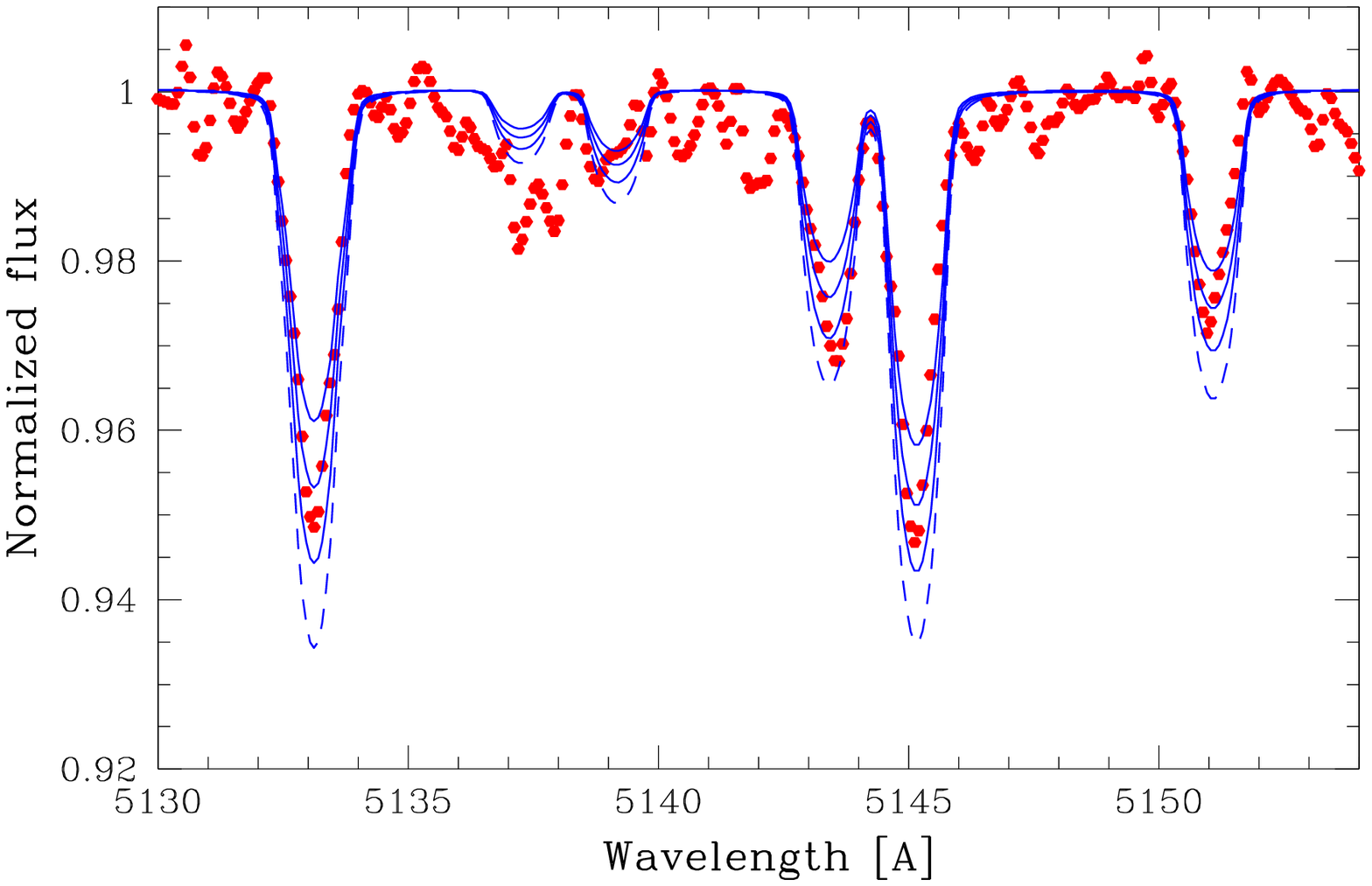}\hspace{5mm}
\includegraphics[width=84mm]{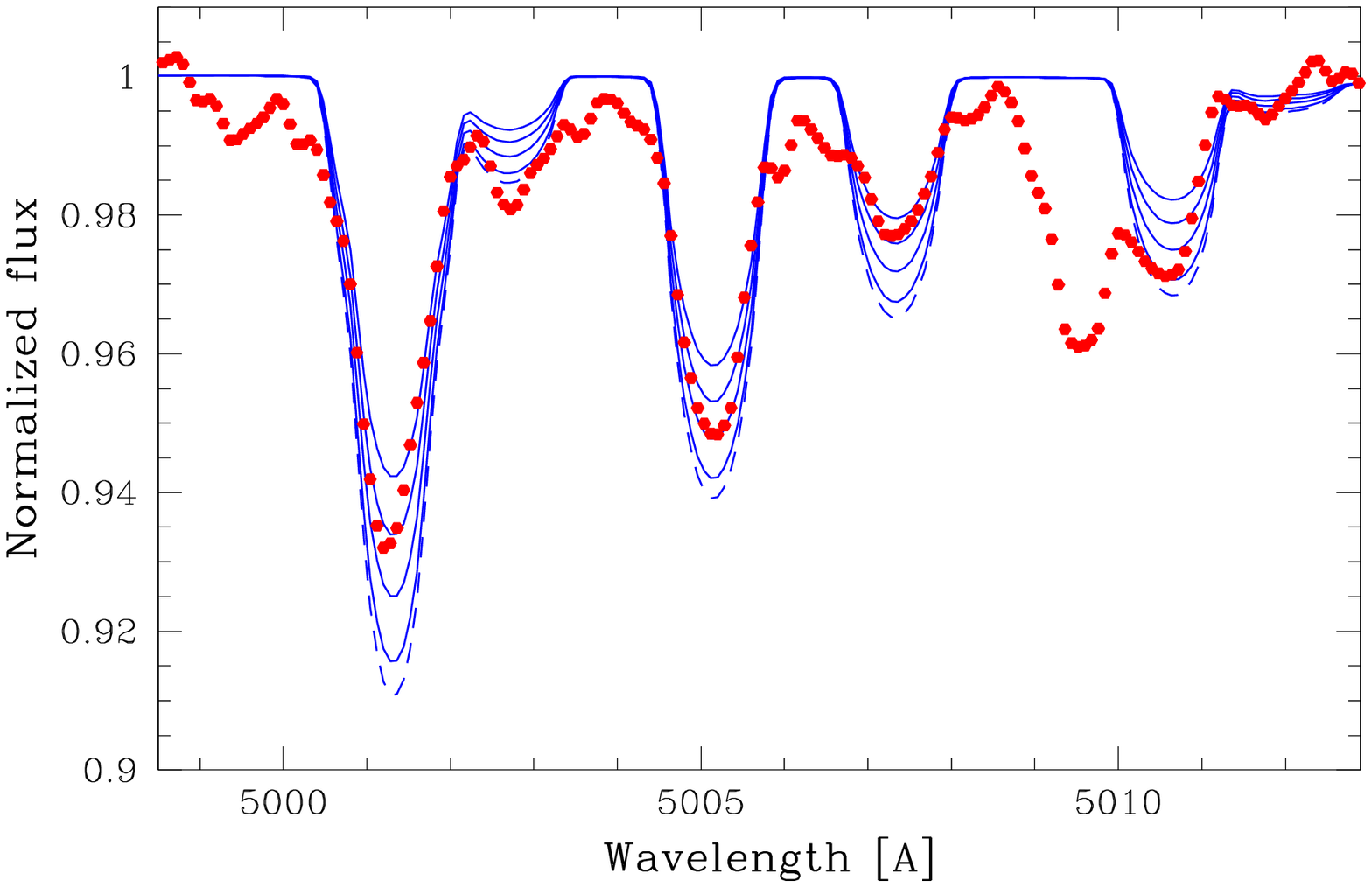}
\caption{\label{Figure: carbon_nitrogen_fits} Comparison between the
disentangled spectra (symbols) of the primary (top) and secondary
(bottom) components with the synthetic spectra (lines) computed
assuming different abundances of C and N. The left and right panels
show bunches of C~{\small I} and N~{\small I} spectral lines,
respectively. The carbon ($\log \epsilon$(C)) and nitrogen ($\log
\epsilon$(N)) abundances range from 8.0 to 8.3~dex, and from 7.4 to
7.6~dex, respectively. The step width is equal to 0.1~dex in both
cases. Dashed lines correspond to the ``present-day cosmic
abundances'' as reported by \citet{Nieva2012}. A colour version of
the figure is available online only.}
\end{figure*}

Abundances of other species present in the spectra of both
components of the V380~Cyg system are given in Table~\ref{Table:
metal_abundances}. The last column lists the difference between the
abundances of the two components. No statistically significant
deviations are found for any of the elements, with oxygen  and
silicon showing the largest discrepancies ($\Delta \log \epsilon =
0.09\pm0.14$~dex and 0.16$\pm$0.20~dex, respectively). Thus, we
conclude that the two stars have a similar chemical composition or
that the difference is smaller than the quoted error bars.
Figure~\ref{Figure: carbon_nitrogen_fits} shows fits to some
individual carbon and nitrogen lines found in the spectra of both
binary components. The synthetic spectra are those used previously
for the determination of the fundamental parameters (see above) but
computed for different abundances of C and N: $\log \epsilon$(C) =
8.0--8.3~dex and $\log \epsilon$(N) = 7.4--7.6~dex. The dashed line
represents the cosmic abundances derived by \citet{Nieva2012}, which
were recently found by \citet{Lyubimkov2013} to be a good
representation of the C and N abundances in some 20 local
($d\leq$\,600~pc from the Sun) B-type main-sequence stars.
Obviously, the derived abundances for both components disagree with
those reported by \citet{Nieva2012} and \citet{Lyubimkov2013}.
However, a large spread in CNO abundances for OB stars was measured
(see an overview in \citealt{Morel2009}). Our abundances agree very
well with the average composition of B-type stars found in the
compilation of available data by this author. Finally, the derived
abundances agree very well with those reported by P2009 for the
primary component of V380\,Cyg, as well as with the abundances found
for other spectral type B members of binary systems: V578\,Mon
\citep{Pavlovski2005} and V453\,Cyg
\citep{Pavlovski&Southworth2009}.

\subsection{Frequency analysis of residual spectra}

\begin{table} \centering
\caption{\label{Table: helium_abundances}The photospheric helium
abundance for both components of the V380 Cyg system as derived from
different He~{\small I} spectral lines.}
\begin{tabular}{lccc}
\hline
Line   &  Primary          &    Secondary    & $\Delta \epsilon$     \\
\hline
4026.2   &  0.087$\pm$0.004  & 0.096$\pm$0.005 & -0.009$\pm$0.006      \\
4387.9   &  0.062$\pm$0.004  & 0.105$\pm$0.004 & -0.043$\pm$0.006      \\
4437.6   &  0.085$\pm$0.003  & 0.081$\pm$0.004 &  0.004$\pm$0.005      \\
4471.5   &  0.094$\pm$0.006  & 0.093$\pm$0.004 &  0.001$\pm$0.007      \\
4713.2   &  0.061$\pm$0.003  & 0.070$\pm$0.004 & -0.009$\pm$0.005      \\
4921.9   &  0.072$\pm$0.004  & 0.089$\pm$0.004 & -0.017$\pm$0.006      \\
5015.7   &  0.079$\pm$0.005  & 0.098$\pm$0.005 & -0.019$\pm$0.007      \\
5054.7   &  0.077$\pm$0.005  & 0.085$\pm$0.005 & -0.008$\pm$0.007      \\
5876.7   &  0.149$\pm$0.013  & 0.096$\pm$0.006 &  0.053$\pm$0.014      \\
6678.1   &  0.083$\pm$0.005  & 0.108$\pm$0.005 & -0.025$\pm$0.007      \\
\hline
mean   &  0.085$\pm$0.019  & 0.092$\pm$0.015 & -0.007$\pm$0.024      \\
\hline
\end{tabular}
\end{table}

\begin{table} \tabcolsep 2.4mm
\caption{\label{Table: metal_abundances} Same as Table~\ref{Table:
helium_abundances} but for metals. Abundances are expressed relative
to $\log N(\rm H) = 12.0$; N gives the number of used lines.}
\begin{tabular}{lcrcrcc}
\hline
Ion      &  Component A  & N   &   Component B  & N   & $\Delta \epsilon $\\
\hline
He       & 10.97$\pm$0.05 & 10 &  11.00$\pm$0.03 & 10  & -0.03$\pm$0.06\\
C        & 8.20$\pm$0.02 & 6   &  8.19$\pm$0.04  & 7   &  0.01$\pm$0.04\\
N        & 7.55$\pm$0.05 & 17  &  7.53$\pm$0.08  & 19  &  0.02$\pm$0.09\\
O        & 8.64$\pm$0.07 & 20  &  8.55$\pm$0.12  & 24  &  0.09$\pm$0.14\\
Mg       & 7.58$\pm$0.10 & 3   &  7.63$\pm$0.11  & 4   & -0.05$\pm$0.15\\
Si       & 7.45$\pm$0.13 & 7   &  7.29$\pm$0.15  & 9   &  0.16$\pm$0.20\\
Al       & 6.02$\pm$0.11 & 3   &  5.98$\pm$0.12  & 3 &
0.04$\pm$0.16\\\hline
N/C      &  -0.65$\pm$0.05 &   &  -0.66$\pm$0.09 &     &  0.01$\pm$0.10\\
N/O      &  -1.09$\pm$0.09 &   &  -1.02$\pm$0.15 &     & -0.07$\pm$0.16\\
\hline
\end{tabular}
\end{table}

The individual RV shifts as well as the mean disentangled spectrum
of the secondary delivered by the spectral disentangling were used
to extract the lines of the primary from the original composite
spectra. First, the contribution of the secondary has been
subtracted from the composite spectra and the residual profiles were
then shifted according to the individual shifts to correct for the
orbital motion. Moreover, in order to prevent the secondary from
influencing our results, similar to the {\sc spd} analysis, we
additionally excluded all eclipse spectra as they may suffer from
the Rossiter-McLaughlin effect \citep{Rossiter1924,McLaughlin1924}
which serves as an additional ``noise source'' for the frequency
analysis. The final set of 237 spectra was used to compute RVs as
the first order moment of the line profile as defined by
\citet{Aerts1992}, which together with the profiles themselves were
the subject of the frequency analysis.

\begin{figure*}
\includegraphics[scale=0.85]{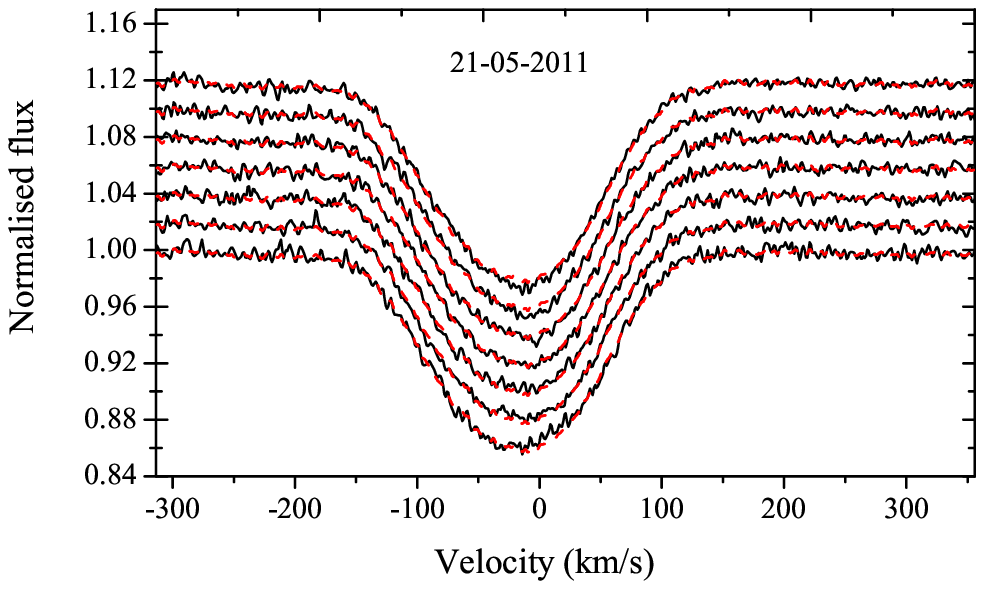}
\includegraphics[scale=0.85]{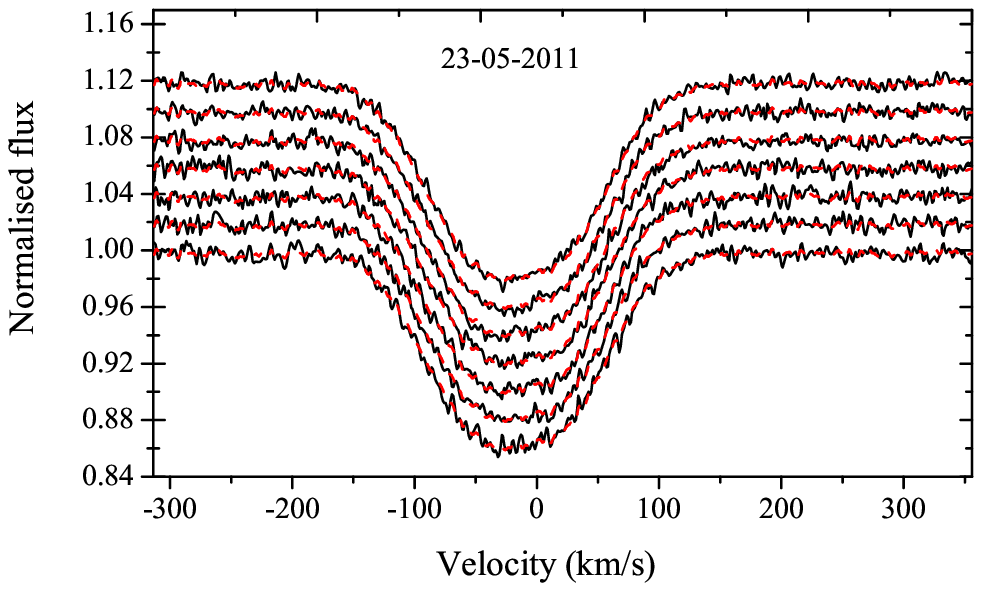}\vspace{3mm}
\centering\includegraphics[scale=0.85]{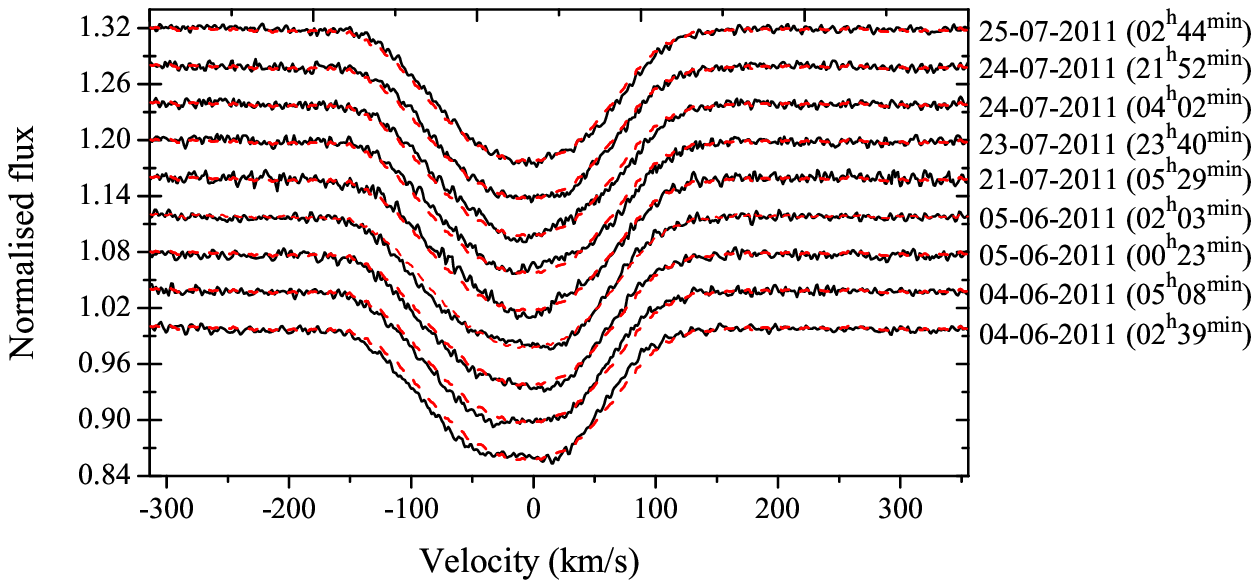}
\caption{\small Line profile variation (LPV) for the primary of
V380\,Cyg after removing the secondary component and correcting for
the orbital motion. {\it Top:} Time-series of spectroscopic
observations during a single night (indicated in the plot). {\it
Bottom:} Nine selected spectra covering about a two months period
from June to July 2011. Mean spectrum (red, dashed line) is given
for comparison for better visibility of the LPV. The value of
\vsini\ of the star equals to 98$\pm$2~\kms\ (cf.
Section~4.2.)}\label{Figure: V380 Cyg - LPV}
\end{figure*}

Figure~\ref{Figure: V380 Cyg - LPV} illustrates time-series of
Si~{\small III} line profiles collected during two single nights
(top panel) as well as nine spectra that cover a much longer period
of time of about two months (bottom). One can see that the spectra
from individual nights do not show any remarkable variation whereas
a clear variability is present on a longer time scale.

For the extraction of the individual frequencies from both RVs and
line profiles, we used the discrete Fourier-transform (DFT) and the
consecutive prewhitening procedure as implemented in the {\sc
famias} package \citep{Zima2008}. The DFT was computed up to the
Nyquist frequency of the data set, and similar to the results of the
photometric analysis, no significant contribution was found in the
high-frequency domain. At each step of the prewhitening, we
optimized the amplitudes and phases of the individual peaks whereas
the frequency values were kept fixed. To make sure that no
significant contribution was missed in the course of our analysis,
we continued extracting frequencies until a S/N of 3.0 was reached.
However, following the definition of the significance level proposed
by \citet{Breger1993} for ground-based data, we consider the
frequency to be significant if it has a S/N$\geq$4.

In total, we analysed six individual spectral lines: three silicon
lines, two lines of helium, and one magnesium line. The results of
the frequency analysis are summarised in the right part of
Table~\ref{Table:Frequency analysis}, labelled ``Spectroscopy''. The
significant frequencies are highlighted in boldface, and each frequency
is linked to the photometric one which in turn is indicated in
brackets. Before we start discussing the individual frequencies
in more detail, it is worth noting that different lines of the same
chemical element delivered compatible results with respect to the
significant frequencies. Since our further analysis concerns only
these frequencies, in Table~\ref{Table:Frequency analysis}, for
silicon and helium, we present the results obtained respectively
from the lines at $\lambda\lambda$~4452~\AA\ and
$\lambda\lambda$~4026~\AA\ only.

Obviously, both pixel-by-pixel and moment methods provide consistent
results for the lines of all the three chemical elements, in the
sense that the signal is detected at the orbital frequency and its
low-order harmonics. Indeed, the orbital signal is the only
significant one in the first moment of all the analysed lines as
well as in the magnesium and helium line profiles. Besides the
orbital frequency and its first two harmonics, the pixel-by-pixel
based analysis of the silicon line profiles reveals a long-term
variability with a frequency of 0.0114~\cd\ (0.1319~\mhz). A similar
frequency of 0.0107~\cd\ (0.1238~\mhz) was also detected in the
photometric data but was found to have very low amplitude (f$_{36}$,
first column of Table~\ref{Table:Frequency analysis}). Multiplied by
a factor of 11, this frequency gives a value of 0.1254~\cd\
(1.4509~\mhz) which is very close to the second, not related to the
binary orbit, significant frequency of 0.1261~\cd\ (1.4590~\mhz)
detected in the silicon line profiles. Moreover, the latter
frequency in turn agrees within the measurement errors with the
rotation frequency of the primary, which we estimated in Section~3.2
to be $\sim$0.1250~\cd\ (1.4463~\mhz). We thus interpret this signal
as being due to the stellar rotation, and suggest that the signal is
modulated with a period of about 90 days (corresponding to the
frequency of 0.0114~\cd\ or 0.1319~\mhz) which might be evidence for
differential rotation. The fact that the rotation signal was
detected in silicon lines only, suggests an inhomogeneous
distribution of the abundance of this element over the stellar
surface. In the next Section, we verify this hypothesis by applying
the Doppler Imaging (DI) method to our spectroscopic data.

\subsection{Doppler Imaging}

The Doppler Imaging technique allows to invert time-series of
observations into the stellar surface distribution maps of different
physical parameters (elemental abundances, temperatures, magnetic
field, etc.). This method is widely used for studying chemically
peculiar (CP) stars that show a non-uniform distribution of one or
several chemical element abundances on their surfaces.

Though stellar surface inhomogeneities were mainly detected in A-
and late B-type as well as in cooler stars so far, we recently
started to observe a similar behaviour in more massive, and hotter
early B-type stars too. Similar to A-type stars, stellar surface
inhomogeneities observed in these objects are usually accompanied by
strong and, sometimes, complex magnetic fields suggesting that it is
this field that is responsible for the formation of spots. For
example, \citet{Kochukhov2011} presented the results of the analysis
of early B-type CP star HD\,37776. The authors reported on the
detection of strong and complex magnetic field as well as on the
inhomogeneous distribution of He on the surface of this star.
\citet{Rivinius2013} reported on strong LPV of both carbon and
helium photospheric lines in the helium strong B star HR\,7355. The
authors attribute the variability to the surface abundance spots
with a rather large, a couple of dex, gradient between the regions
of enhancement and depletion of the chemical element.
\citet{Degroote2011} and \citet{Papics2012} reported on the
detection of the rotational signal from both CoRoT photometry and
high-resolution spectroscopy in two B-type stars HD\,174648 and
HD\,43317, respectively. \citet{Degroote2011} showed that for
HD\,174648, the signal can be explained by the two circumpolar
spots, whereas the spot-like scenario for HD\,43317 is supported by
the recent detection of a magnetic field for this star
\citep{Briquet2013}.
\citet{Petit2013} presented a summary of physical, rotational,
and magnetic properties of all known magnetic O and B stars (see
their Table~1). \citet{Hubrig2006} reported the detection of a weak
longitudinal magnetic field of the order of a few hundred of Gauss
in 13 SPB stars and one $\beta$\,Cep variable ($\xi^1$\,CMa)
although this was recently criticized
\citep{Bagnulo2012,Shultz2012}.

\begin{table*} \tabcolsep 1.8mm\caption{\small Results of the frequency analysis. The Rayleigh limit amounts to 0.0013~\cd\ for the photometry and 0.0071~\cd\ for the spectroscopy. Errors in the amplitude are given in parenthesis in terms of last
digits. Frequencies matching the significance criterion of
S/N$\geq$4.0 are highlighted in boldface. Each spectroscopic
frequency is linked to the photometric one. The corresponding
photometric frequency is indicated in parenthesis following the
value of the spectroscopically derived frequency.}
\begin{tabular}{lllll|llll|llll} \hline
\multicolumn{5}{c|}{Photometry\rule{0pt}{9pt}} &
\multicolumn{8}{c}{Spectroscopy}\\
\hline
 & & & & & \multicolumn{4}{c|}{Pixel-by-pixel\rule{0pt}{9pt}} & \multicolumn{4}{c}{RV}\\
\multicolumn{5}{c|}{(magnitude)} & \multicolumn{4}{c|}{(continuum
units)} & \multicolumn{4}{c}{(\kms)}\\ \hline f$_i$ & Freq (\cd) &
Amplitude & S/N & n$\cdot$f$_{\rm orb}$ & Freq\rule{0pt}{9pt} (\cd)
&
Amplitude & S/N & n$\cdot$f$_{\rm orb}$ & Freq (\cd) & Amplitude & S/N & n$\cdot$f$_{\rm orb}$\\
\hline
f$_1$ & {\bf 0.08054}\rule{0pt}{9pt} & 0.001204(215) & 5.0 & 1.00 & \multicolumn{8}{c}{Silicon lines ($\lambda\lambda$ 4552.6, 4567.8, 4574.8\,\AA)\rule{0pt}{9pt}}\\
f$_2$ & {\bf 0.12946} (f$_{\rm rot}$) & 0.001284(208) & 5.1 & 1.61 & {\bf 0.0114} (f$_{36}$) & 0.0401(26) & 4.1 & 0.14 & {\bf 0.0822} (f$_{1}$) & 0.9900(352) & 5.3 & 1.02\\
f$_3$ & {\bf 0.25879} (2f$_{\rm rot}$) & 0.001131(192) & 5.1 & 3.21
& {\bf 0.0822} (f$_{1}$) & 0.0090(20) & 5.6 & 1.02 & 0.2425
(f$_{52}$)& 0.7767(372) & 3.6 & 3.01\\\cline{1-5}
f$_7$\rule{0pt}{9pt} & 0.14161 & 0.000829(186) & 3.6 & & {\bf 0.2426} (f$_{52}$) & 0.0078(21) & 4.1 & 3.01 & {\bf 0.1630} (f$_{18}$)& 0.6678(393) & 4.7 & 2.02\\
f$_{17}$ & 0.22832 & 0.000606(161) & 3.1 & & {\bf 0.1606} (f$_{18}$) & 0.0075(22) & 4.0 & 1.99 & 2.1014 & 0.3511(358) & 3.7 & \\
f$_{18}$ & 0.16474 & 0.000624(159) & 3.0 & & 0.6397 (f$_{50}$)& 0.0061(20) & 3.4 & & 0.8916 (f$_{99}$)& 0.3853(382)& 3.8 & \\
f$_{25}$ & 0.29562 & 0.000589(148) & 2.9 & & {\bf 0.1261} (f$_{2}$) & 0.0063(21) & 4.0 & 1.57 & 0.6058 (f$_{165}$)& 0.2268(352) & 3.5 &\\
f$_{35}$ & 0.44763 & 0.000469(132) & 2.8 & & 0.2909 (f$_{25}$) & 0.0058(21) & 3.1 & & & & &\\
f$_{36}$ & 0.01067 & 0.000664(132) & 2.8 & & 0.2345 (f$_{17}$)& 0.0052(20)& 3.0 & & & & &\\
f$_{50}$ & 0.64702 & 0.000412(117) & 2.8 & & \multicolumn{8}{c}{Helium lines ($\lambda\lambda$ 4026.6, 4471.5\,\AA)}\\
f$_{51}$ & 0.48624 & 0.000351(115) & 2.8 & & {\bf 0.0792} (f$_{1}$) & 0.0126(19) & 5.3 & 0.98 & {\bf 0.0796} (f$_{1}$) & 0.8702(400)& 5.3 & 0.99\\
f$_{52}$ & 0.24531 & 0.000409(114) & 2.7 & & 0.4461 (f$_{35}$) & 0.0076(19) & 3.6 & & 0.4467 (f$_{35}$)& 0.2861(409) & 3.7 &\\
f$_{61}$ & 0.44192 & 0.000355(108) & 2.6 & & 0.8488 (f$_{67}$) & 0.0075(20) & 3.8 & & 0.8497 (f$_{67}$) & 0.3546(399) & 3.4 &\\
f$_{67}$ & 0.85214 & 0.000323(105) & 2.8 & & 0.1462 (f$_{7}$)  & 0.0071(21) & 3.1 & & 0.2425 (f$_{52}$) & 0.4285(456) & 3.1 &\\
f$_{99}$ & 0.89541 & 0.000266(88)  & 2.6 & & \multicolumn{8}{c}{Magnesium line ($\lambda\lambda$ 4481.1\,\AA)}\\
f$_{131}$ & 0.97831 & 0.000241(76) & 2.6 & & 0.9829 (f$_{131}$) & 0.0138(12) & 3.7 & & {\bf 0.0805} (f$_{1}$) & 1.0214(504) & 5.9 & 1.00\\
f$_{165}$ & 0.60610 & 0.000192(65) & 2.6 & & {\bf 0.0793} (f$_{1}$) & 0.0051(11) & 4.9 & 0.99 & 0.4461 (f$_{35}$) & 0.5414(503) & 3.8 &\\
f$_{216}$ & 0.98671 & 0.000147(53) & 2.5 & & 0.9896 (f$_{216}$) & 0.0045(12) & 3.2 & & 1.2163 (f$_{281}$) & 0.5046(522) & 3.6 &\\
f$_{281}$ & 1.21676 & 0.000123(43) & 2.6 & & 0.4461 (f$_{35}$) & 0.0025(11) & 3.0 & & 0.2934 (f$_{25}$) & 0.3932(546) & 3.3 &\\
\hline
\end{tabular}
\label{Table:Frequency analysis}
\end{table*}

\begin{figure*}
\includegraphics[angle=90,scale=0.40,clip=]{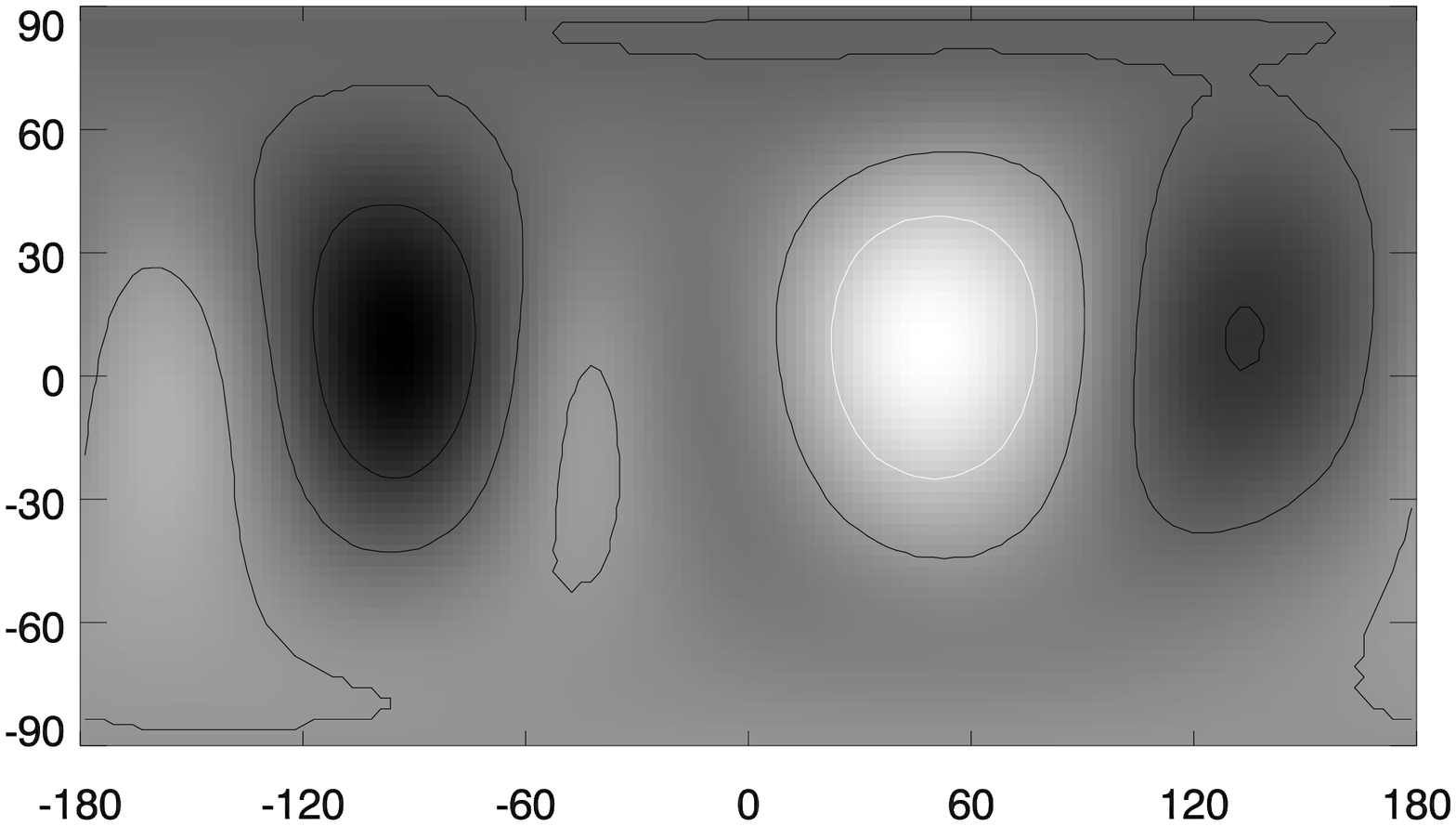}\hspace{3mm}
\includegraphics[scale=0.85,clip=]{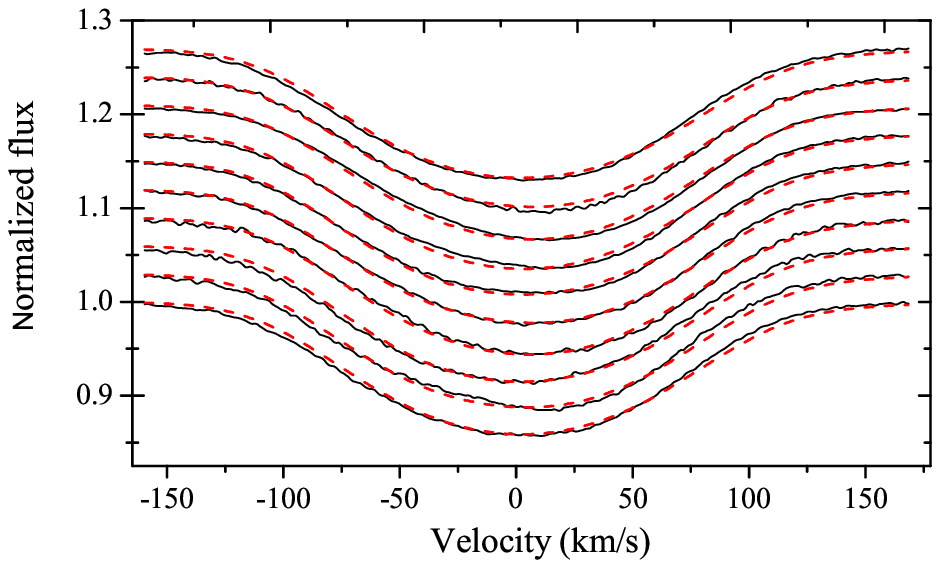}\vspace{2mm}
\includegraphics[angle=90,scale=0.40,clip=]{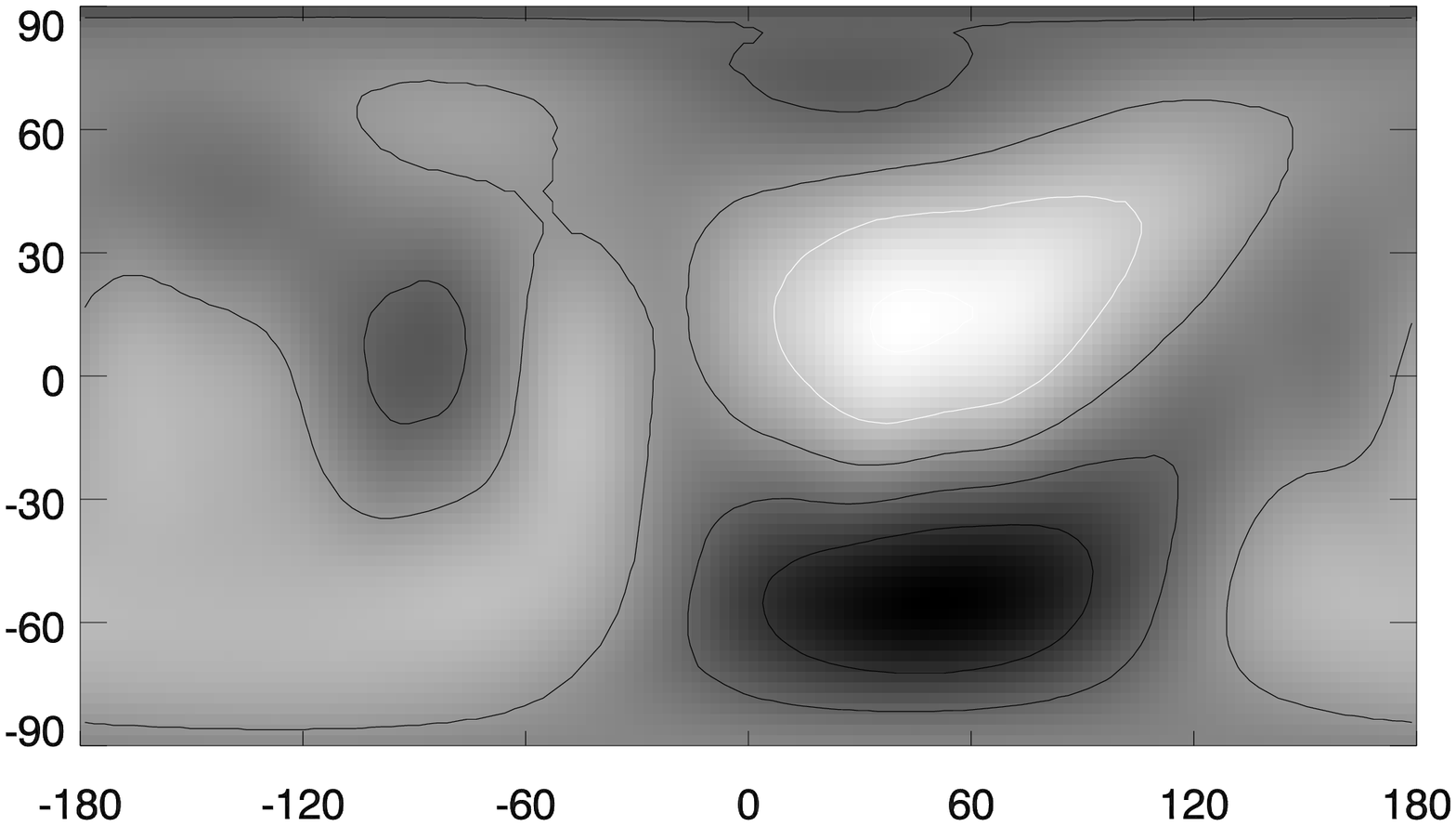}\hspace{3mm}
\includegraphics[scale=0.85,clip=]{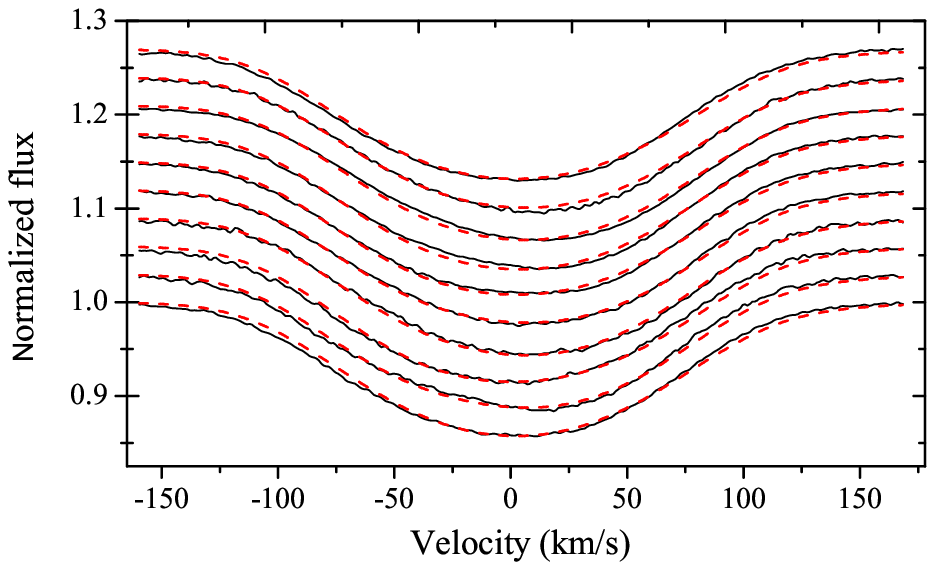}\vspace{2mm}
\includegraphics[angle=90,scale=0.40,clip=]{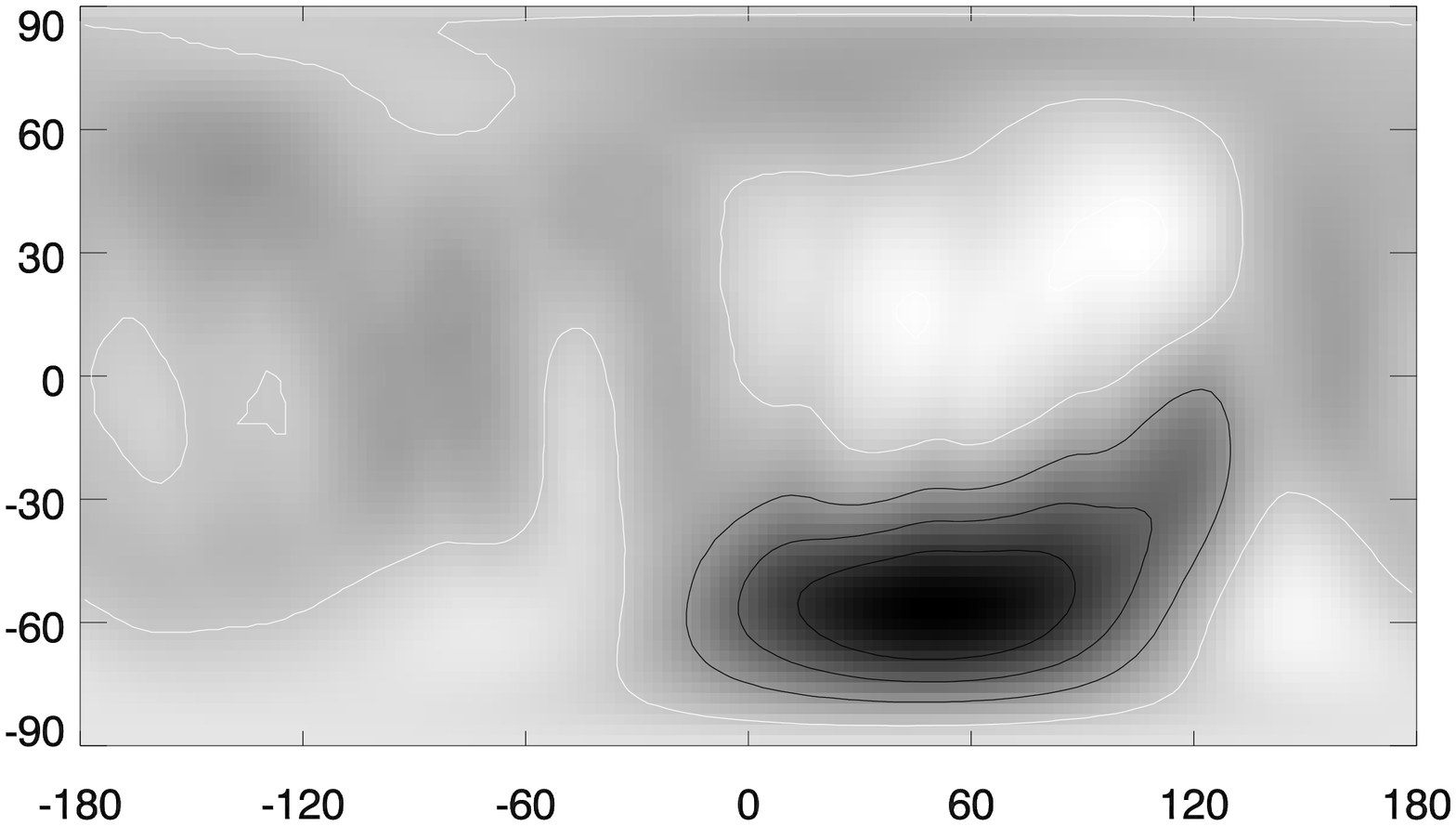}\hspace{3mm}
\includegraphics[scale=0.85,clip=]{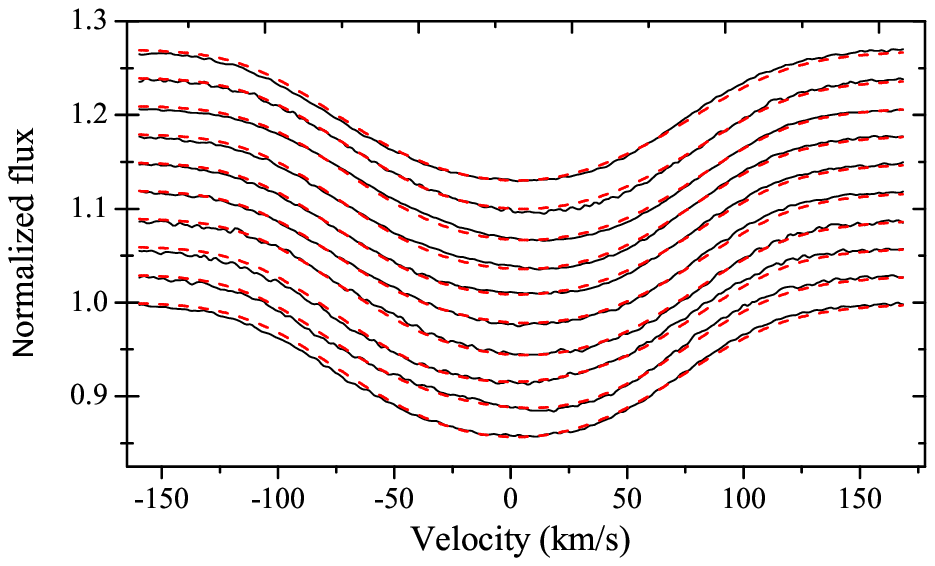}
\caption{\small Si-abundance surface maps in Mercator projection
(left) and the comparison between the observed (solid, black line)
and computed (dashed, red line) profiles. The calculations are based
on three values of the regularization parameter $\Lambda$:
10$^{-2}$, 10$^{-3}$, and 10$^{-4}$ (from top to bottom). The
profiles corresponding to different rotation phases are vertically
shifted by a small, constant value for better visibility. X- and
Y-axes on the maps give longitude $\theta$ and latitude $\phi$
coordinates, respectively. The dark and bright areas refer to the
regions of enhanced and depleted Si abundance, respectively. The
maximal abundance gradient increases from the top map to the bottom
and is (in dex): [Min:Max]=[-4.41:-4.10] -- top, [-4.65:-3.61] --
middle, and [-4.69:-2.56] -- bottom.}\label{Figure: Mercator maps}
\end{figure*}

There are several theories to explain the existence of the magnetic
fields in massive, early-type stars. The most disseminated theory is
the one of fossil magnetic fields, i.e., the remnant of the magnetic
field gained by the star in the course of its formation phase
\citep{Braithwaite2006}. There are also theories proposing that the
convective core of the massive stars \citep{Brun2005} or a dynamo
process operating in the differentially-rotating radiative layers
\citep{Mullan2005} is at the origin of the magnetic fields observed
at the surfaces of these stars. \citet{Cantiello2009,Cantiello2011}
suggested, from their theoretical calculations and the MHD
simulations, the presence of a sub-surface convective region in
massive stars that is capable of producing localized magnetic
fields observable at their surfaces. The authors also suggested that
this convective region might be responsible for the large
microturbulent velocity fields typical for massive early-type stars.

The fact that Si~{\small III} line profiles of the primary component
of V380\,Cyg show a clear variability with the period close to the
(expected) rotation period of the star, suggests that the system may
contain yet another rotationally modulated B-type star. Phase-folded
silicon profiles on which we base our further analysis (see below),
show the same kind of variability as shown in Figure~\ref{Figure:
V380 Cyg - LPV} (bottom).

We used the Si~{\small III} spectral line at
$\lambda\lambda$~4452~\AA\ to check whether the observed LPV can be
explained by the rotation effect due to spot(s) on the stellar
surface. The analysis is based on the above described method of DI
as implemented in the {\sc invers8} Fortran code
\citep{Piskunov1993}. The code inverts time-series of the observed
line profiles into surface abundance distribution maps by minimizing
the difference between the observations and the model. Since the
problem is strongly ill-posed, it has to be regularized which is
done by means of Tikhonov regularization. From a mathematical point
of view, the regularization is performed by adding an additional
term of the form $\Lambda{\rm F(\epsilon)}$ to the total error
function that is a subject of minimization, where $\Lambda$
represents a regularization parameter (usually, taking a value
between 10$^{-2}$ and 10$^{-4}$) and F($\epsilon$) stands for the
regularization functional. The algorithm then iteratively reduces
the total error function until the discrepancy between the
observations and the model is of the same order as the individual
errors of measurement \citep{Kochukhov2003}.

Calculations with the {\sc invers8} code are based on the
pre-computed tables of the local line profiles. These profiles are
computed for different positions on the stellar disc (20 in our
case), which allows to take the centre-to-limb variation of the
intensity, and for a set of individual abundances of a given
chemical element. The profiles are then convolved with the
instrumental profile of a certain width, which in our case
corresponds to the high-resolution mode (R=85\,000) of the {\sc
hermes} instrument \citep{Raskin2011}. Since the primary component
of V380\,Cyg is a hot B-type giant star, non-LTE line formation must
be assumed when synthesizing its line profiles. The local profiles
were computed with the non-LTE version of the SynthV code developed
by VT. To synthesize the spectra, the code uses pre-computed non-LTE
departure coefficients, which in our case were calculated with the
{\sc tlusty} model atmospheres code \citep{Hubeny1988,Hubeny1995}.

In our calculations with the {\sc invers8} code, we limited the
total number of iterations to 500 but in all cases the program
converged much faster, typically after completing 170-200
iterations. The calculations were performed using three values of
the regularization parameter $\Lambda$: 10$^{-2}$, 10$^{-3}$, and
10$^{-4}$. The resulting Si-abundance maps for all the three values
of $\Lambda$ are shown in Figure~\ref{Figure: Mercator maps} (left)
in Mercator projection. Obviously, in all three cases, the algorithm
favours a solution with two ``dominant'' spots, of which one is
bright (abundance depletion) and another is dark (abundance
enrichment). The map computed for $\Lambda$=10$^{-2}$ also exhibits
a third spot at a longitude of about 130$^{\circ}$ but it is of much
lower contrast compared to the bulk of the star than the two others.
The essential difference between this map (top) and the two other
maps (middle and bottom) is that the spots are located at almost the
same latitude $\phi$ but at different longitudes $\theta$ such that
only one of the high-contrast spots is visible at a given instant of
time (except for the phases when both spots are visible at the very
limb of the star). In the two other cases, the spots have rather the
same longitude but different latitudes which makes them invisible
for almost half of the rotation cycle. The quality of the fit of the
observed profiles for all three maps is almost the same, as is
illustrated in Figure~\ref{Figure: Mercator maps} (right). The major
part of the observed spectroscopic variability can be explained by
either of our models, also confirming the suggestion made in
Sections~3.2 and 4.4 that the star shows a signature of rotational
modulation.

A careful look at the obtained maps and the corresponding abundance
gradients allows us to exclude the one computed for the lowest value
of $\Lambda$=10$^{-4}$ (cf. Figure~\ref{Figure: Mercator maps},
bottom). This map exhibits an unreliably high gradient of the Si
abundance of about 2~dex between the two spots, which is compensated
by other small-scale structures on the stellar surface (e.g.,
low-contrast but elongated spots at [$\theta$,
$\phi$]=[-135$^\circ$, +55$^\circ$]; [-75$^\circ$, +0$^\circ$]) to
provide a good fit to the observations. Moreover, the rather high
mean abundance (without taking the two spots into account) of
$\sim$3.5~dex characteristic of this map is not compatible with the
Si abundance estimated from the spectrum analysis of the
disentangled spectrum of the primary (cf. Section~4.2). To some
extent, this also concerns the map computed for $\Lambda$=10$^{-3}$
(cf. Figure~\ref{Figure: Mercator maps}, middle), as the estimated
mean Si abundance for this map is $\sim$3.95~dex which also seems to
be too high. We thus favour the map showing two large spots located
at almost the same latitude (cf. Figure~\ref{Figure: Mercator maps},
top) though we are aware that it is rather difficult to decide which
of the two maps ($\Lambda$=10$^{-2}$ or 10$^{-3}$) is the most
appropriate one.

\begin{figure}
\includegraphics[angle=90,scale=0.36,clip=]{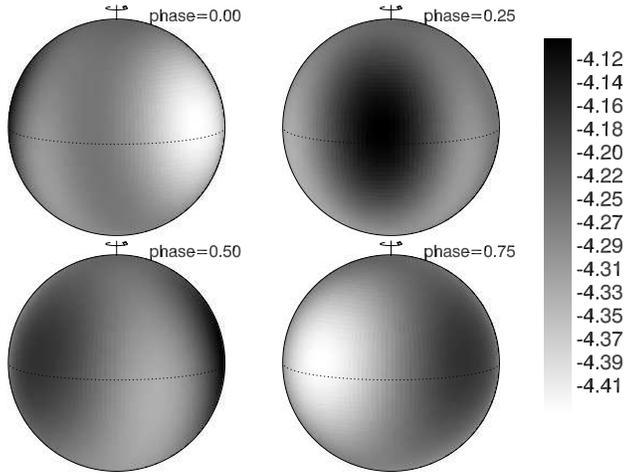}
\caption{\small Si-abundance map in spherical projection computed
for $\Lambda$=10$^{-2}$. The dark and bright areas refer to the
regions of enhanced and depleted Si abundance,
respectively.}\label{Figure: Spherical maps}
\end{figure}

Figure~\ref{Figure: Spherical maps} shows the selected map in
spherical projection and at four different rotational phases. Though
the detected total Si-abundance gradient of about 0.3~dex is small,
it is significant compared to the estimated errors. As stated by
\citet{Piskunov1993}, the most reliable errors on the reconstructed
abundances are obtained by ``mapping'' several spectral lines of the
same chemical element separately. We thus performed the DI analysis
of all the three Si lines used by us for the frequency analysis (cf.
Table~\ref{Table:Frequency analysis}). The surface abundance
distributions obtained from the individual lines are compatible with
each other, also when assuming different values of the
regularization parameters $\Lambda$. The delivered abundance
gradients are consistent as well, giving a maximum abundance
amplitude of $\sim$0.35~dex for the map computed with
$\Lambda$=10$^{-2}$. The errors estimated from the abundance scatter
between the maps obtained based on the three individual spectral
lines are of the order of 0.1~dex, thanks to the very high S/N of
our spectroscopic data, which makes the detection of such a low
abundance gradient possible. The continuum normalization of the
spectra usually also serves as an additional source of uncertainty
on the reconstructed maps, but is minimal in our case due to the
high-quality of the data.

As an additional check for the consistency between the frequency
analysis of the line profiles and the DI results, we applied the
algorithm to both He and Mg lines in which we could not detect the
signal that we interpret as being due to rotation of the star.
Similar to the case of silicon, we used the same spectral lines as
for the frequency analysis, i.e., two He and one Mg lines (cf.
Table~\ref{Table:Frequency analysis}). It is worth noting that a
visual inspection of the lines of both elements folded with the
rotation period, was enough to conclude that they do not exhibit
variability similar to the silicon lines but are stable over the
rotation cycle instead. Indeed, the DI algorithm delivers rather
structured surface maps even for the regularization parameter of
$\Lambda$=10$^{-2}$ and with a very low abundance gradient below
0.1~dex. Such a gradient is considered insignificant as it is
comparable to the estimated uncertainties, and we conclude that both
helium and magnesium show a uniform distribution of their abundances
over the stellar surface.

\subsection{Evolutionary models}

As already mentioned (cf. Section~1), both components of the
V380\,Cyg system show a discrepancy between the dynamical mass and
the one obtained from evolutionary models. P2009 showed that the
discrepancy in mass does not disappear with the inclusion of
rotation effects into the evolutionary models, whereas G2000
suggested a high value of the convective overshoot parameter
($\alpha_{ov}=0.6\pm0.1$) as a possible solution for the mass
problem. In the present study, the fundamental stellar parameters of
both components of V380\,Cyg are refined, and used to compare to a
new set of evolutionary model calculations. We use the MESA stellar
evolution code \citep{Paxton2011,Paxton2013} to compute the tracks.
The basic setup includes an initial hydrogen mass fraction of
$X=0.70$ and metal abundance fractions from \citet{Asplund2009}. The
initial rotation velocity on the pre-main-sequence track is set to
$v_\mathrm{init}=0$~\kms\ (no rotation), and we choose to include
core overshooting as a step function, parameterized with a
$\alpha_\mathrm{ov}=0.2$ pressure scale height by default. The
Schwarzschild criterion is used in the convection treatment. We
choose a standard mixing length $\alpha_\mathrm{MLT}=1.8$ pressure
scale height. The MESA equation-of-state is used, and the N$^{14}$
reaction rate from \citet{Imbriani2004} We compared the stellar
models computed with MESA to those in the grid of stellar models
computed with the Code Li\'egois d'\'Evolution Stellaire (CLES) by
\citet{Briquet2011} and used by \citet{Saesen2013} for asteroseismic
modelling of B-type pulsators in the open cluster NGC\,884.
Excellent agreement was found for the same input physics.

\begin{figure*}
\includegraphics[width=82mm]{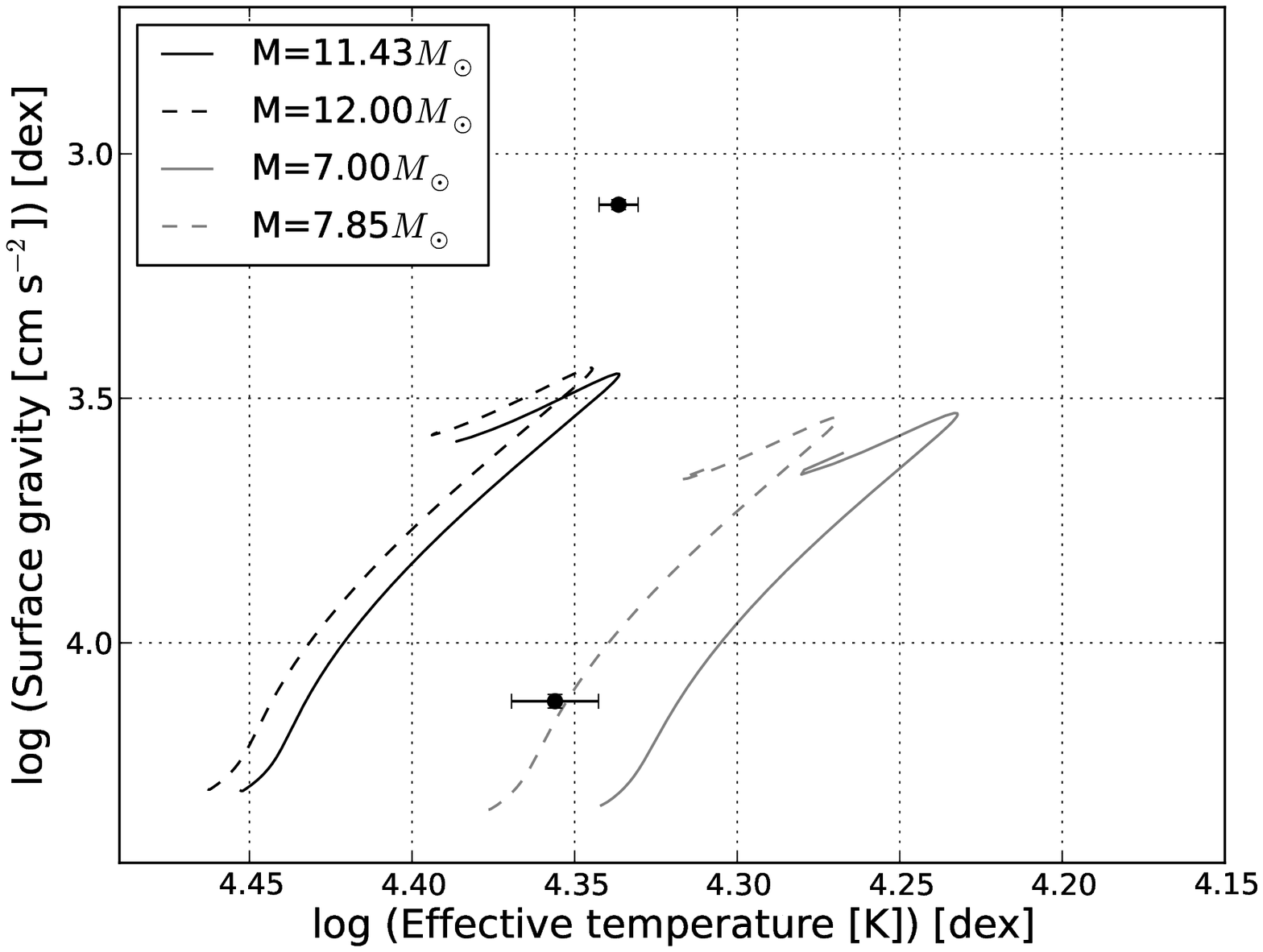}\hspace{5mm}
\includegraphics[width=82mm]{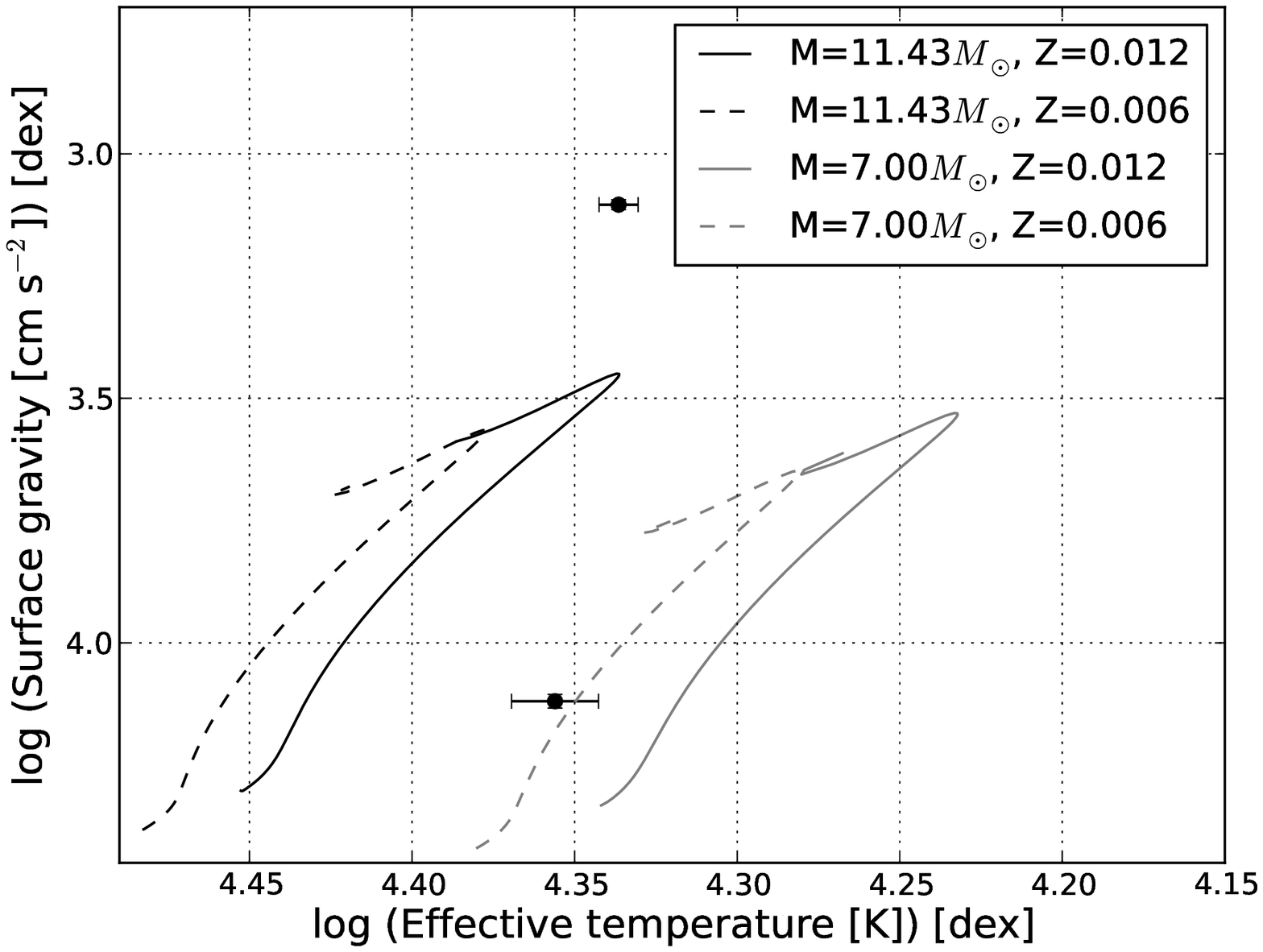}
\caption{\label{Figure: HR_diagram} Location of the primary and
secondary components of V380~Cyg in the \te-\logg\ diagram.
Theoretical evolutionary tracks show the influence of mass (left)
and metallicity (right).}
\end{figure*}

\begin{figure*}
\includegraphics[scale=0.47]{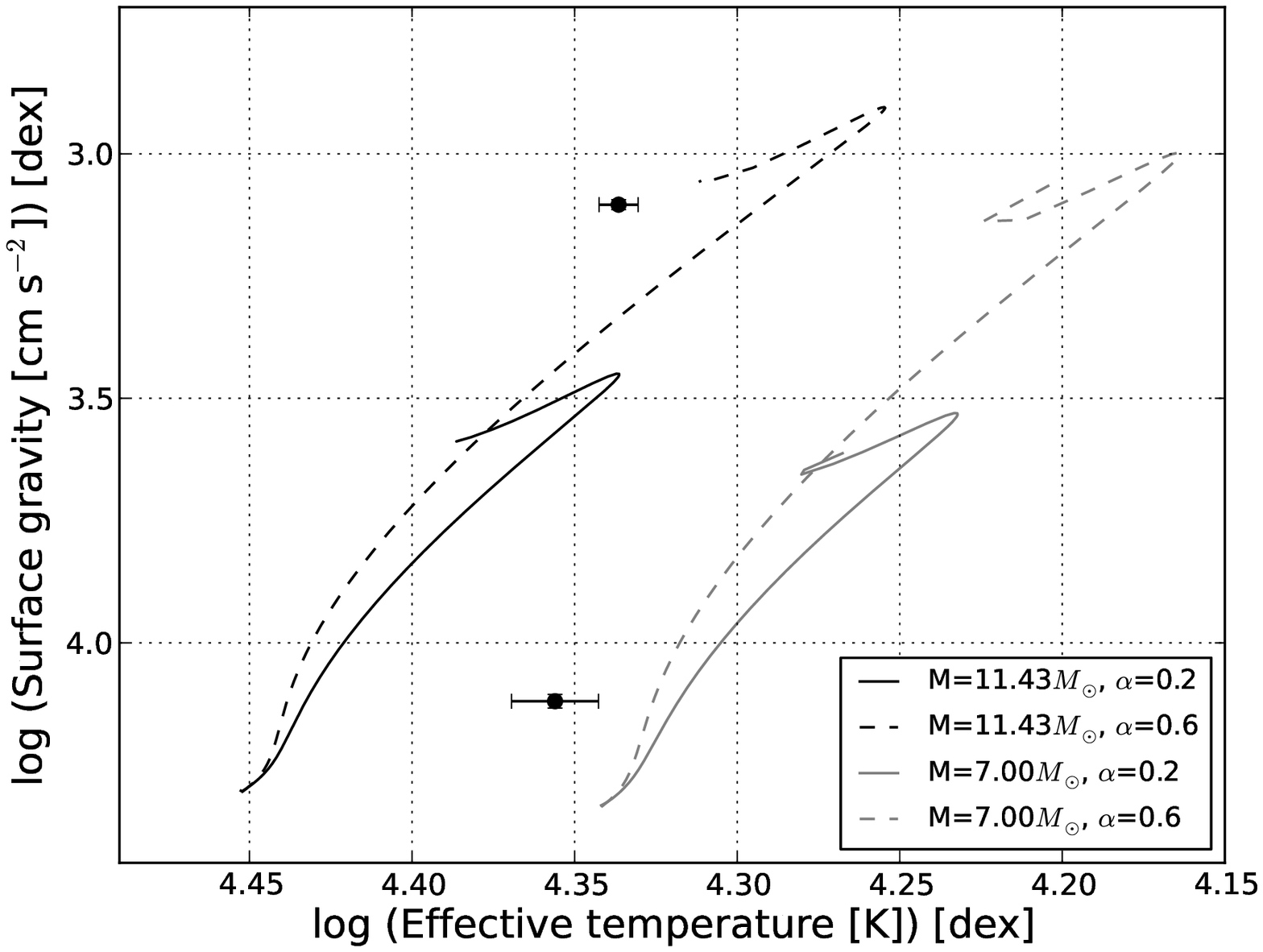}\hspace{5mm}
\includegraphics[scale=0.47]{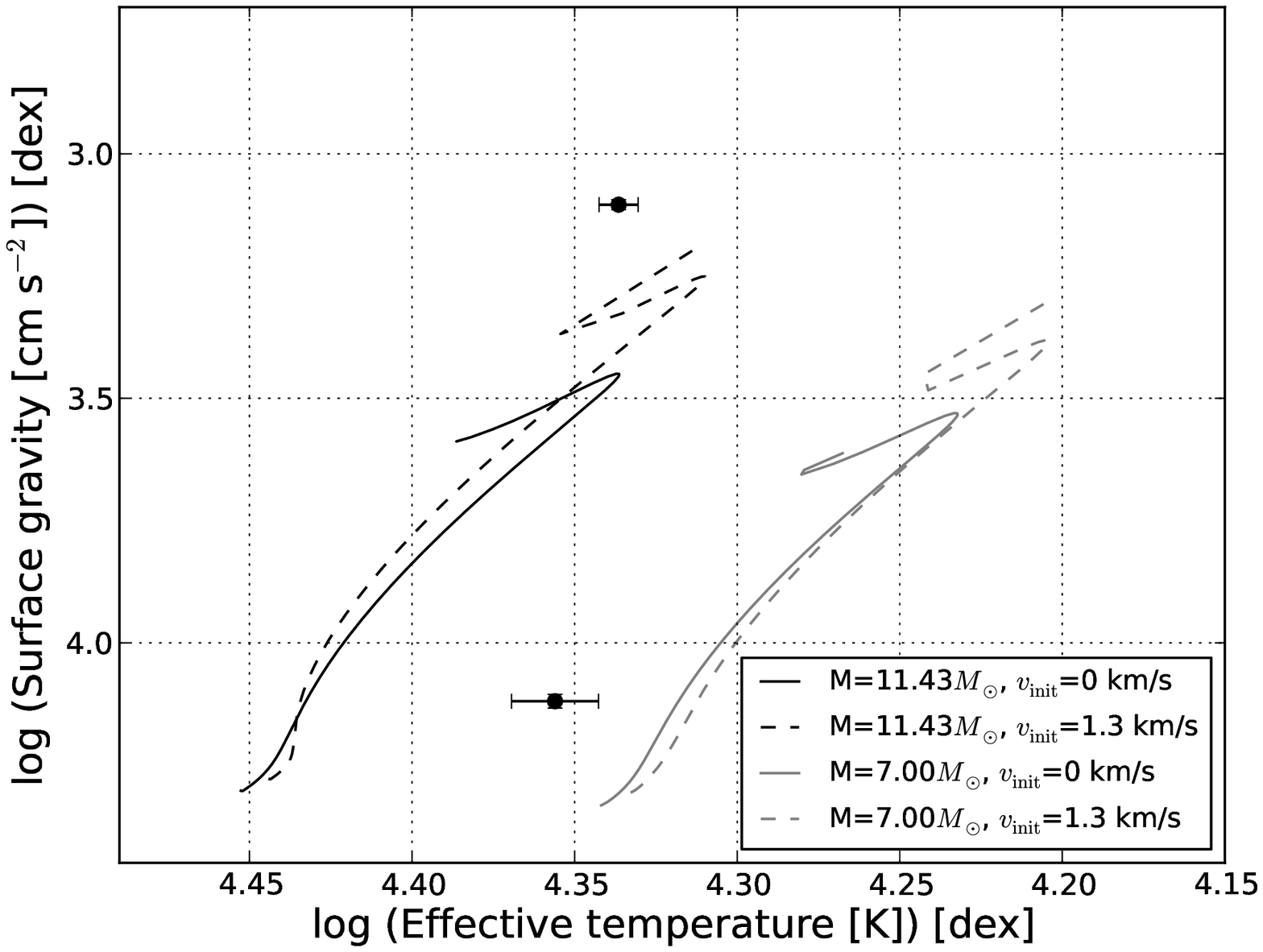}
\caption{\label{Figure: HR_diagram1} Same as Fig.~\ref{Figure:
HR_diagram} but for the influence of overshooting (left) and initial
rotational velocity (right).}
\end{figure*}

\begin{figure}
\includegraphics[scale=0.47]{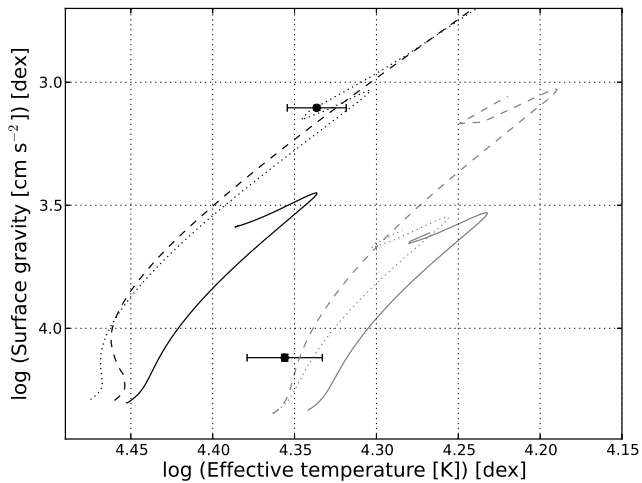}\hspace{5mm}
\caption{\label{Figure: HR_diagram2} Location of the primary and
secondary components of V380~Cyg in the \te-\logg\ diagram. Solid,
long- and short-dashed tracks correspond to model 1, 2, and 3 in
Table~\ref{Table:models}, respectively.}
\end{figure}

Figures~\ref{Figure: HR_diagram}-\ref{Figure: HR_diagram2} show the
location of both components of V380\,Cyg in the \te-\logg\ diagram
along with the evolutionary tracks. Figure~\ref{Figure: HR_diagram}
illustrates the influence of mass and bulk metallicity on the
theoretical tracks. One can see that both parameters have a similar
effect on the models, namely a mass increase or metallicity decrease
shifts the tracks to the left in the \te-\logg\ diagram, i.e.,
towards higher effective temperatures. The effects of core
convective overshooting and of the initial rotational velocity are
shown in Figure~\ref{Figure: HR_diagram1}. We see that both
phenomena increase the stellar life time on the main-sequence, with
the effect being larger for the overshooting.

From Figure~\ref{Figure: HR_diagram} (left panel), it is clear that
a discrepancy of about 0.85$M_{\odot}$ between the dynamical and
theoretical masses exists for the secondary component. The right
panel of Figure~\ref{Figure: HR_diagram} shows that one can match
these two masses by assuming a lower bulk metallicity for the star
($Z=0.006$). The model calculations represent the case of a
non-rotating star and assume a value of $\alpha_\mathrm{ov}=0.2$. A
non-rotating model appears to be reasonable approximation for the
secondary since its estimated rotation rate is 8.0$\pm$0.8\% of the
critical velocity. However, the assumed low bulk metallicity for the
secondary is not consistent with the analysis of its disentangled
spectrum, which showed close to solar atmospheric abundances (cf.
Section~4.2). The two models that fit the position of the star in
the \te-\logg\ diagram, taking into account the error bars, are
illustrated in Figure~\ref{Figure: HR_diagram2} by long- and
short-dashed lines (hereafter models 2 and 3, respectively). Both
models assume the same mass of 7.42~$M_{\odot}$ (within 3-$\sigma$
from the dynamical mass derived in Section~3.1), chemical
composition $Z=0.012$~dex, and no rotation, but a significantly
different overshooting parameter of $\alpha_\mathrm{ov}=0.6$ and 0.2
for models 2 and 3, respectively. The estimated age of the star from
these two models is 21.5$\pm$1.5 and 18$\pm$1~Myr, respectively.

Assuming that both components of V380 Cyg have the same (initial)
metal content ($Z=0.012$) and age, and that the primary could be
either in the core or shell hydrogen burning phase of the evolution
(models 2 and 3 in Table~\ref{Table:models}, respectively), we
attempted to fit its position in the \te-\logg\ diagram. As we can
see in Figure~\ref{Figure: HR_diagram2}, for model 2 to match the
observations, we need to introduce an unreasonably high rotation
($v_\mathrm{ZAMS}=240$~\kms which corresponds to the rotational
velocity of 330~\kms\ at the location of the star in
Figure~\ref{Figure: HR_diagram2}), an extremely strong amount of
overshooting ($\alpha_\mathrm{ov}=0.6)$, and a mass at the $3\sigma$
limit ($M=12.00$~$M_{\odot}$). For comparison with the models, we
choose the error bars on the effective temperature of the primary to
be at the 3-$\sigma$ level of the spectroscopic estimate of 300~K
(cf. Table~\ref{Table: atmospheric_parameters}). The latter value is
the standard deviation of the mean which hardly represents realistic
errors (cf. Section~4.2). We find the uncertainty of 900~K to be
more conservative, which allows (at least partly) to take into
account the effects of non-uniform temperature distribution on the
surface of the primary due to its non-spherical shape (the star
rotates at $\sim$32\% of its critical velocity), imperfect continuum
normalization and atomic data, etc. We also show a model 3 that
allows to fit the position of the primary in the \te-\logg\ diagram
assuming the star is in the shell hydrogen burning phase
(short-dashed line in Figure~\ref{Figure: HR_diagram2}). Although
this model assumes $\alpha_\mathrm{ov}=0.3$, we still had to
introduce an unreasonably high rotation ($v_\mathrm{ZAMS}=243$~\kms
which corresponds to the rotational velocity of 420~\kms\ at the
location of the star in the diagram) as well as to increase mass of
the star by $\sim$1.5~$M_{\odot}$ compared to the dynamical value.
Increasing the mass of the primary enables us to match the observed
location of the primary with track prediction, but it is not
reconcilable with the radial velocity measurements. The MESA model
predictions clearly point to mass discrepancy for the primary
component, in agreement with the findings by G2000 and P2009.

\section{Discussion and Conclusions}

In this paper, we presented a detailed study of V380\,Cyg, an
eccentric binary ($e = 0.22$) with two massive components (11.4 and
7.0~M$_{\odot}$ for the primary and secondary, respectively) of
spectral type B. We based our analysis on high-precision photometric
data gathered with the {\em Kepler} space-based telescope, as well
as on ground-based high-dispersion spectroscopy obtained with the
{\sc hermes} spectrograph \citep{Raskin2011}. The data were analysed
using state-of-the art methods to determine orbital parameters of
the system and absolute dimensions of both stellar components, as
well as to explain both photometric and spectroscopic variability
intrinsic to the primary star and discovered only recently by
\citet{Tkachenko2012}.

The effects of binarity in the light curve were modelled using the
Wilson-Devinney code \citep{Wilson1971,Wilson1979}, while the
spectral disentangling technique was used to derive the
spectroscopic orbit as well as the disentangled spectra of both
stellar components. The determined orbital and physical parameters
agree well with those reported by \citet{Guinan2000} and
\citet{Tkachenko2012}. However, we find a considerable difference
between the temperature of the secondary obtained from the light
curve fitting ($T_{\rm eff,B} = 23\,840\pm500$~K) and the one
deduced from the spectral characteristics of this star ($T_{\rm
eff,B} = 21\,800\pm400$~K). We find a similar chemical composition
for both binary components and a good agreement with the abundances
reported by \citet{Pavlovski2009} for the primary component of
V380\,Cyg.

After subtracting the orbital signal from both the photometric and
spectroscopic data, the residuals were subjected to frequency
analysis. The results of the photometric and spectroscopic analyses
are consistent with each other in the sense that the variability
occurs at the orbital frequency and its harmonics as well as at the
rotation frequency of the primary. All other variability detected in
the {\em Kepler} data is of stochastic nature. We find that the
characteristics of this variability are not in contradiction with
the recent theoretical predictions by \citet{Cantiello2009} and
\citet{Shiode2013} for the g-mode oscillations excited both in the
core and in a thin convective sub-surface layer of massive stars.

We also explored the hypothesis of observing the rotationally
modulated signal in our data in more detail. We used the Doppler
Imaging technique to study variability detected in several spectral
lines of doubly ionized silicon. Two solutions that provide nearly
the same quality of the fit but different spot configurations on the
stellar surface have been obtained. We favour the one that shows a
lower abundance gradient and assumes two high- and one low-contrast
spot located at (nearly) the same latitude but at different
longitudes.


We used the MESA stellar evolution code to look at the model
predictions for both components of V380\,Cyg. For the secondary, we
found a discrepancy of $\sim$0.85~$M_{\odot}$ between the dynamical
and theoretically predicted masses assuming standard stellar
evolutionary physics. However, a slightly larger value of the mass
(within 3-$\sigma$ from the dynamical value) and a bulk metallicity
of $Z=0.012$~dex can account for this difference. Two models that
fit the position of the primary in the \te-\logg\ diagram were found
and provide significantly different age estimates for the system:
21.5$\pm$1.5 and 18$\pm$1~Myr (cf. Table~\ref{Table:models}). In
both cases, the system appears to be much younger than 25.5$\pm$1.5
Myr, the age determined by G2000 assuming $\alpha_\mathrm{ov}=0.6$
for the primary component. Since both models assume extreme input
physics like a very high rotation at the ZAMS and/or a large amount
of overshooting, we conclude that single star evolutionary models
are not suitable for this particular binary system.

The variability detected in the {\em Kepler} data is intrinsic to
the primary component and of stochastic nature. The signal is
variable both in amplitude and in appearance on a short time-scale.
This makes the tuning of the convective core overshooting parameter
$\alpha_{ov}$ from asteroseismic analysis impossible for this star.
In future papers, we plan to take a closer look at another massive
binary system, Spica, with the aim to see whether the above
mentioned mass discrepancy holds for this star too and whether
asteroseismology can help us to constrain $\alpha_{ov}$ parameter in
that case.

\begin{table} \tabcolsep 1.2mm\caption{\small Evolutionary model parameters for both components of V380\,Cyg.}
\begin{tabular}{llllllll} \hline
Par & \multicolumn{1}{c}{Unit\rule{0pt}{9pt}} &
\multicolumn{3}{c}{{\bf Primary}} & \multicolumn{3}{c}{{\bf
Secondary}}\\
& & \multicolumn{1}{c}{Model1} & \multicolumn{1}{c}{Model2} & \multicolumn{1}{c}{Model3} & \multicolumn{1}{c}{Model1} & \multicolumn{1}{c}{Model2} & \multicolumn{1}{c}{Model3}\\
\hline
$M$ & $M_{\odot}$ & 11.43 & 12.00 & 12.90 & 7.00 & 7.42 & 7.42\\
$Z$ & dex & 0.014 & 0.012 & 0.012 & 0.014 & 0.012 & 0.012\\
$\alpha_\mathrm{ov}$ & $Hp$ & 0.2 & 0.6 & 0.3 & 0.2 & 0.6 & 0.2\\
$v$ & \kms & 0 & 241 & 243 & 0 & 0 & 0\\
Age & Myr & --- & 21.5 & 18 & --- & 21.5 & 18\\
\hline
\end{tabular}
\label{Table:models}
\end{table}

\section*{acknowledgements}
The research leading to these results received funding from the
European Research Council under the European Community's Seventh
Framework Programme (FP7/2007--2013)/ERC grant agreement
n$^\circ$227224 (PROSPERITY). This research is (partially) funded by
the Research Council of the KU Leuven under grant agreement
GOA/2013/012. The research leading to these results has received funding from the European Community's Seventh Framework Programme FP7-SPACE-2011-1, project number 312844 (SPACEINN)

\label{lastpage}

\end{document}